\documentclass[preprint]{aastexr}

\newcommand{\ha}{H$\alpha$}

\shorttitle{ALMA quiet Sun}
\shortauthors{Alissandrakis et al.}

\begin{document}

\title{The quiet Sun at mm Wavelengths as Seen by ALMA\footnote{Accepted for publication Frontiers in Astronomy and Space Science}} 

\correspondingauthor{Costas E. Alissandrakis}
\email{calissan@cc.uoi.gr}

\author{Costas E. Alissandrakis}
\affiliation{Deparment of Physics, University of Ioannina, 45110 Ioannina, Greece}

\author{T. S. Bastian}
\affiliation{National Radio Astronomy Observatory, 520 Edgemont Road, Charlottesville, VA 22903 USA}

\author{Roman  Braj{\v{s}}a}
\affiliation{Hvar Observatory, Faculty of Geodesy, University of Zagreb, Ka\u ci\' ceva 26, 10000 Zagreb, Croatia}

\begin{abstract}
\noindent Solar observations at sub-mm, mm and cm wavelengths offer a straightforward diagnostic of physical conditions in the solar atmosphere because they yield measurement of brightness temperature which, for optically thick features, equals intrinsic temperature - much unlike solar diagnostics in other spectral ranges.  The Atacama Large Millimeter and sub-millimeter Array (ALMA) has therefore opened a new, hitherto underexplored, spectral window for studying the enigmatic solar chromosphere.  In this review we discuss initial ALMA studies of the quiet chromosphere that used both single-dish and compact-array interferometric observing modes.  We present results on the temperature structure of the chromosphere, comparison with classic empirical models of the chromosphere, and observations of the chromospheric network and spicules.  Furthermore, we discuss what may be expected in the future, since the ALMA capabilities continuously expand and improve towards higher angular resolution, wavelength coverage, and polarization measurement for magnetometry.
\end{abstract}


\section{Introduction}
\label{sec:Intro}

Traditionally, the ``quiet Sun" refers to the solar atmosphere outside of solar active regions and during times free of transient energetic events such as flares and coronal mass ejections. Despite decades of study at optical (O), ultraviolet (UV), and extreme ultraviolet (EUV) wavelengths, a detailed understanding of the quiet Sun has remained elusive. ALMA has now opened a new window onto the Sun: millimeter- and submillimeter-wavelength observations of the Sun with angular and time resolutions that are orders of magnitude better than was previously possible, comparable to those available at O/UV/EUV wavelengths. The Atacama Large Millimeter and sub-millimeter Array (ALMA, \citet{2009IEEEP..97.1463W}) offers new and complementary diagnostics of solar phenomena, including the quiet Sun. In this review we present and assess ALMA observations of the quiet Sun (QS). We do not treat transient phenomena and oscillations, which are the subject of the review by Nindos et al., in this special Research Topic collection. Comparisons with radiative Magnetohydrodynamics (rMHD) models are addressed in the review by Wedemeyer et al., included in this special Research Topic collection.

We start with a brief description of  O/UV observations of the QS in order to put the mm-$\lambda$ observations in perspective (Section~\ref{quiet}) and then we discuss what ALMA can offer to our understanding of the physical processes involved  (Section~\ref{ALMA?}). We proceed with the presentation of results obtained from low resolution full-disk ALMA observations, on the morphology (Section~\ref{morph}), on empirical models of the solar atmosphere (Section~\ref{Section:CLV}) and on the formation height of the emission (Section~\ref{Sect:height}). The results from high resolution interferometric observations are discussed in Section~\ref{Section:HR}, for disk (Section~\ref{Section:disk}) and limb (Section~\ref{Sect:spicules}) structures, whereas the appearance of limb structures on the disk  is discussed in Section~\ref{Sect:spicdisc}. We conclude with a brief discussion and some thoughts about the prospects of future observations of the quiet Sun with ALMA. 

\section{The quite Sun in optical and EUV wavelengths}\label{quiet}

The solar atmosphere is commonly described in terms of the visible {\sl photosphere}; the {\sl chromosphere}, the thin layer visible in H$\alpha$ during eclipses; and the {\sl corona}, the extended atmosphere of the Sun visible in white light during eclipses. The {\sl transition region} refers to material that lies at the interface between the chromosphere and the corona. Of particular interest here is the chromosphere because it is from this medium that radiation at mm and submm wavelength originates as we now discuss.

\begin{figure}[h]
\begin{center}
\includegraphics[height=7cm]{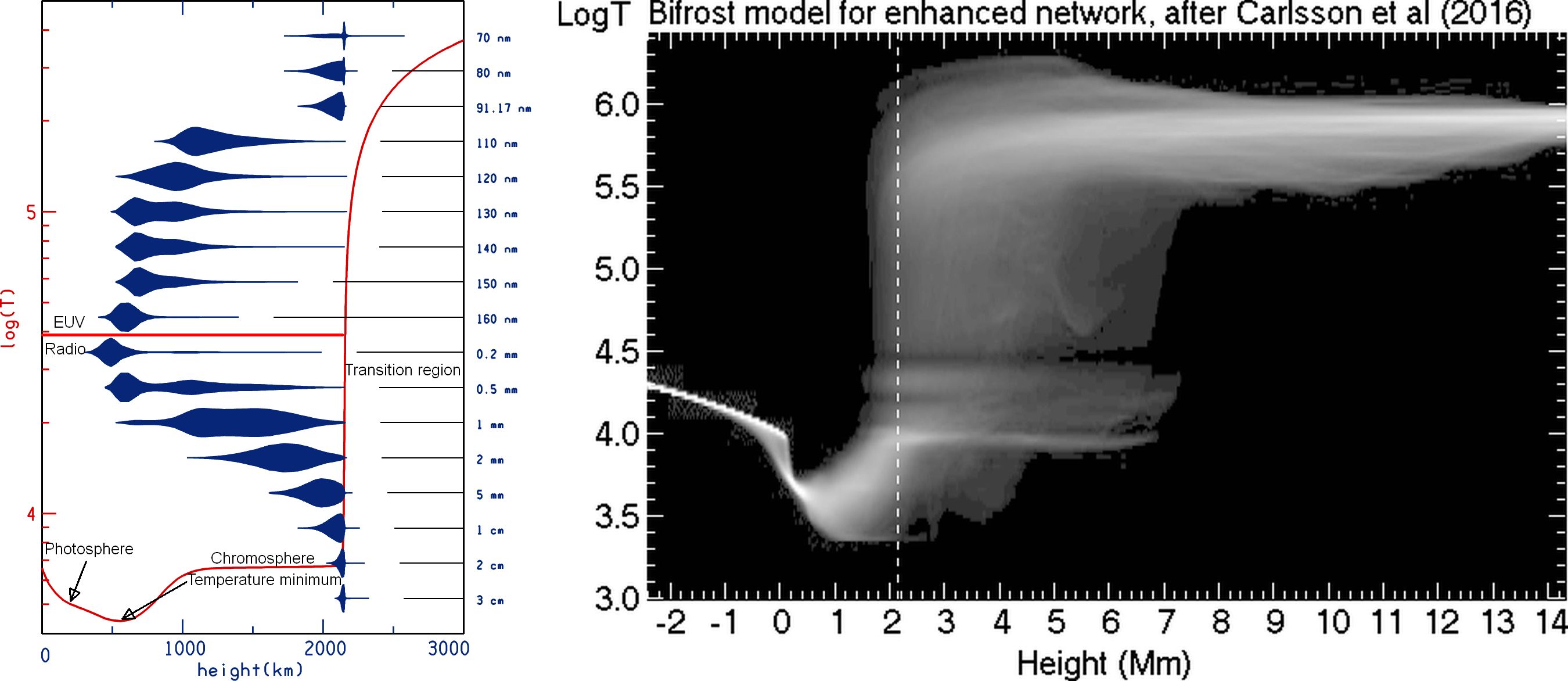}
\end{center}
\caption{The temperature as a function of height and formation height of mm-$\lambda$ emission. {\bf(Left)} The electron temperature as a function of height (red line) according to model C7, adapted from \citet{2008ApJS..175..229A}. The contribution to the continuum intensity is plotted in blue for wavelengths 0.2-30\,mm (below the red horizontal line) and the extreme ultraviolet wavelengths, marked at the right. {\bf(Right)} Probability density function of the temperature as function of height from Bifrost simulations of an enhanced network region. The vertical dashed line is at 2.14\,Mm. Adapted from the lower panel of Figure 10 of \citet{2016A&A...585A...4C}.
}
\label{model}
\end{figure}

As a first approximation, the physical parameters of the solar atmosphere can be assumed to vary with height only. This is because gravity produces a strong stratification and thus the radial density gradient is much larger than the horizontal, at least in the lower atmospheric layers. This assumption has led to the classic one-dimensional atmospheric models, often multi-component to describe various atmospheric features (see the reviews by \citet{2011SoPh..273..309S} and \citet{2020FrASS...7...74A} for more details). The temperature structure of the solar atmosphere according to one such model is plotted in red in the left panel of Figure~\ref{model}. Once the temperature and density structure are known, one can compute the opacity and the effective height of formation of the radiation at a particular wavelength; this is also plotted in the left panel of the same figure, in the form of contribution functions, which give the distribution of the observed emission with height and are equal to the height derivative of the emergent intensity, $dI_\nu/dh$. We note that the solar mm radio emission comes from the chromosphere, possibly extending down to the temperature minimum at sub-mm wavelengths. We further note that extreme ultraviolet (EUV) continuum emission also comes from the same atmospheric layers (upper part of Figure~\ref{model}, top left), thus images in these two spectral ranges are expected to show similar structures; hence, combining mm and EUV continuum data can potentially provide improved diagnostics.

With the advent of fast numerical computations, a number of sophisticated tools have been developed for 3D solar atmospheric modeling (see review by Wedemeyer et al., included in this special Research Topic collection). Such models clearly demonstrate the complex structure and the dynamics of the upper layers of the solar atmosphere (see below). An example of the deduced temperature structure is given in the right panel of Figure~\ref{model}, from a simulation of an enhanced network region by \citet{2016A&A...585A...4C}. It is interesting that transition region temperatures, from various features, appear in an extended range of heights that go beyond 2.14\,Mm (dashed line), predicted by \citet{2008ApJS..175..229A}. Although this particular model may not be representative of the average QS, which includes inter-network regions, it is still indicative.

\begin{figure}[h]
\begin{center}
\includegraphics[width=15cm]{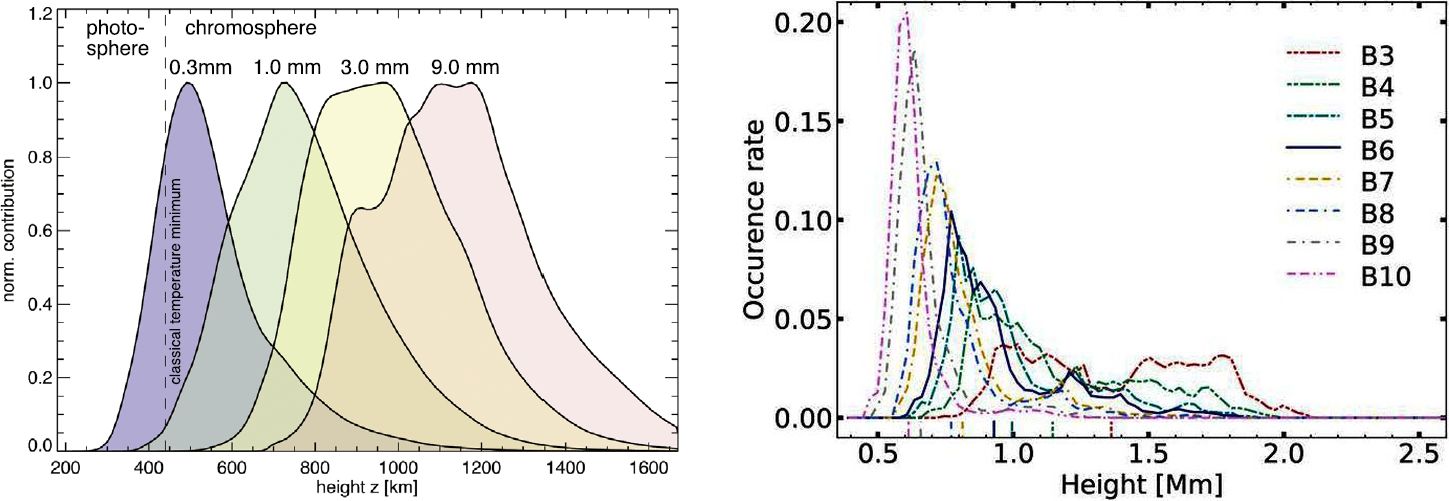}
\end{center}
\caption{Contribution functions for mm-$\lambda$. {\bf(Left)} Contribution functions at four wavelengths between 0.3 mm and 9.0 mm at the center of the solar disk according to \citet{2016SSRv..200....1W}. Reprinted by permission from: Springer Nature, Space Science Reviews, \copyright\ 2019. {\bf(Right)} QS contribution functions for the ALMA frequency bands according to \citet{2021A&A...656A..68E}. Reproduced with permission \copyright\ ESO.
}
\label{model2}
\end{figure}

The average contribution functions computed from such a model at mm-$\lambda$ are displayed in the left panel of Figure~\ref{model2} (see also Figure 3 in Wedemeyer et al., this special Research Topic collection). They are similar to the predictions of the 1D model, putting again the formation height of mm emission above the temperature minimum and into the chromosphere. More recently, numerical simulations under local thermodynamic equilibrium by \citet{2020ApJ...891L...8M} gave average formation heights of 0.9\,Mm (with a standard deviation of 0.7\,Mm) for the 1.2\,mm emission and 1.8\,Mm (standard deviation of 1\,Mm) for the 3\,mm emission, whereas computations by the same authors under non-equilibrium hydrogen ionization gave greater heights and a much smaller height difference: $2.67\pm1.08$\,Mm for 1.2\,mm emission and $2.78\pm1.09$\,Mm for 3\,mm emission. Finally, \citet{2021A&A...656A..68E} computed contribution functions for all ALMA frequency bands (Figure~\ref{model2}, right) and predicted lower heights: 0.9\,Mm for the QS at 1.2\,mm and 1.3\,Mm at 3\,mm. Observational evidence on the formation height is discussed in Section~\ref{Sect:height}.

Of course, anyone who has seen a solar images knows that horizontal structure is very important. Briefly, horizontal structure in the solar atmosphere comes from the interplay of two factors: one of them is mass motions and the other is the magnetic field. Magnetohydrodynamics (MHD), which is a good (but not perfect) approximation for the solar atmosphere, tells us that the result of the interaction depends on the energy density of each factor: when the magnetic energy density (which is the same as the magnetic pressure) is much larger than the energy density of the plasma (which includes thermal and kinetic energy or, alternatively gas and dynamic pressure), the plasma flows along the magnetic field lines of force (see, e.g., \citet{2020FrASS...7...74A}, also \citet{2019ARA&A..57..189C}, for a more detailed discussion). This is the case in the chromosphere and the corona, where we see spectacular structures that map the magnetic field, as well as in sunspots.\footnote{This comparison is often made in terms of the plasma parameter $\beta$, which is the ratio of the gas pressure to the magnetic pressure.} 

As we go down to the photosphere, the density increases faster than the magnetic field \citet{2001SoPh..203...71G}, and thus the plasma dominates in that atmospheric layer. Here convective motions on two scales, the {\it granulation} (see, e.g., Figure 5 in \citet{2020FrASS...7...74A}) and the larger {\it supergranulation} determine the horizontal structure, with the latter pushing and compressing the magnetic field at the borders of the supergranules, thus forming the {\it chromospheric network}. This particular cellular structure, although detectable in photospheric velocity maps, manifests itself mainly in the chromosphere, hence its name.

\begin{figure}[h]
\begin{center}
\includegraphics[width=\textwidth]{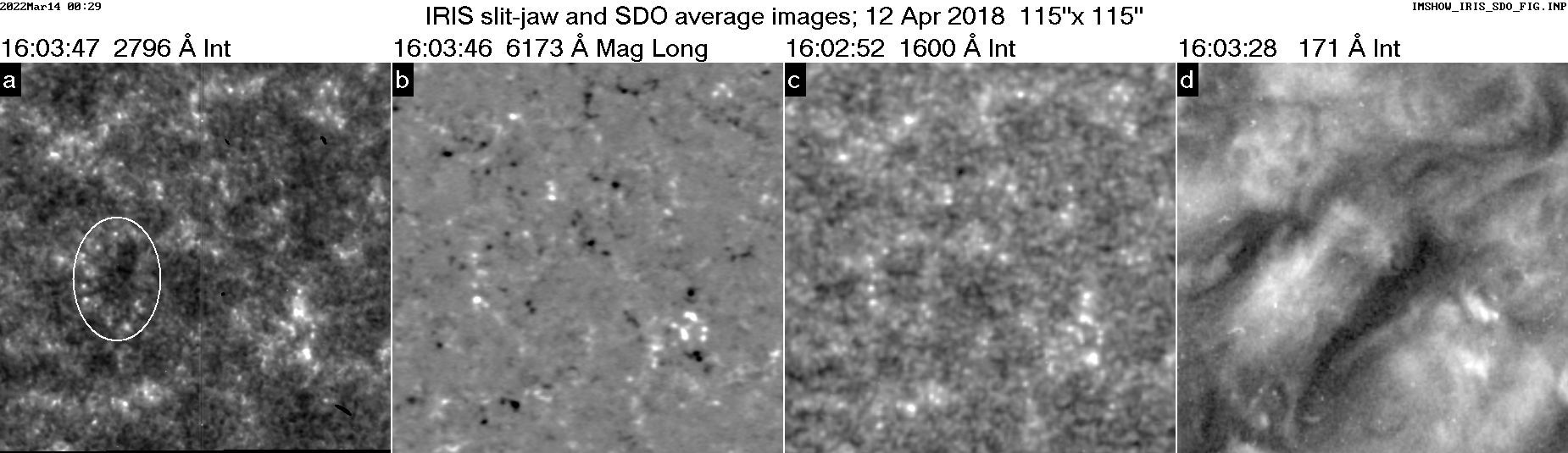}
\end{center}
\caption{Images of a quiet Sun region: (a) in the Mg\,{\sc ii} k line; (b) an HMI magnetogram, saturated at $\pm50$\,G; (c) in the AIA 1600\,\AA\ band and (d) in the AIA 171\,\AA\ band. Images have been averaged over 20\,min, to reduce the effect of 5-minute oscillations which is very prominent at 2796 and 1600\,\AA. The white ellipse in panel a marks a prominent supergranulation cell. Figure made by the authors from IRIS, AIA and HMI data. 
}
\label{network}
\end{figure}

At the chromospheric level, the network appears in the form of bright structures, well visible in the core of the Ca{\sc II} K line, known as {\it bright mottles} (coarse or fine depending on the spatial resolution of the instrument; see the reviews by \citet{1968SoPh....3..367B}, \citet{1972ARA&A..10...73B} and \citet{Tsiropoula2012}). It is also visible in all chromospheric spectral lines and continua, such as the Mg\,{\sc ii} k line at 2796\,\AA, observed by the {\it Interface Region Imaging Spectrograph} (IRIS) and the UV continua centered at 1600 and 1700\,\AA\ spectral bands observed by the {\it Atmospheric Imaging Assembly} (AIA), aboard the {\it Solar Dynamics Observatory} (SDO). Figure~\ref{network} shows such images, together with a magnetogram from the {\it Helioseismic and Magnetic Imager} (HMI) aboard SDO and an AIA image in the 171\,\AA\ band. We note the cellular arrangement of the network structures and their association with regions of enhanced magnetic field. We also note that the structures disappear in the 171\,\AA\ image, illustrating the well-know fact that the network becomes diffuse in the transition region and disappears in the low corona \citep{1974ApJ...188L..27R}; apparently, this is a result of fanning out of the magnetic field, and of field lines closing at low heights.

In addition to the bright mottles, \ha\ images show absorbing features dubbed as {\it dark mottles}\footnote{Some authors use the term {\it fibrils} instead of mottles; however, classically, fibrils are elongated structures seen in active regions where the magnetic field has a strong horizontal component.}. These structures appear on top of the bright mottles and are best visible in the blue wing of the \ha\ line, suggesting ascending motions (see top panel of Figure 9 in \citet{2020FrASS...7...74A}).  As the quality of the observations improves, new results and new names appear in the literature; for example, \cite{2008ApJ...679L.167L} reported dark features on the solar dik, in the blue wing of the Ca ii IR line, and so did \citet{2009ApJ...705..272R} for same line and H$\alpha$; these {\it rapid blueshifted events} were identified as the counterpart of type II spicules (see below). Dark mottles are hard to see in other chromospheric lines or continua (see, however, \citet{2019A&A...631L...5B, 2021A&A...647A.147B}), probably because they are geometrically and optically thin. 

\begin{figure}%[h]
\begin{center}
\includegraphics[width=\textwidth]{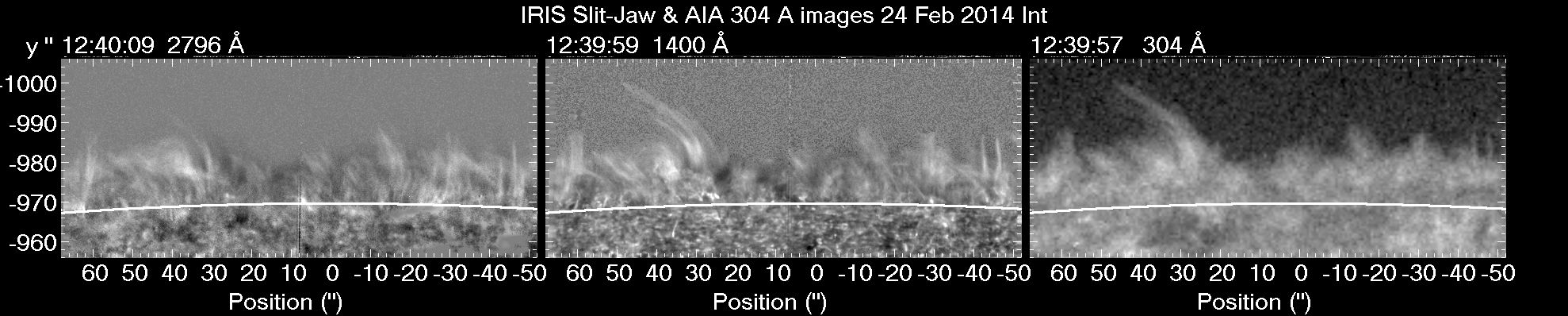}
\end{center}
\caption{Spicules beyond the limb near the south pole of the sun, as seen in IRIS slit-jaw images with a bandwidth of 4\,\AA\ around the Mg{\sc\,ii} 2796\,\AA\ k line (left), and with a spectral width of 55\,\AA\ in the 1400\,\AA\ band which includes the Si{\sc\,iv} lines (center), as well as in an AIA image in the He{\sc\.ii} 304\,\AA\ line (right). Time-averaged images have been subtracted from the original IRIS data for better visibility of spicules. The white arc marks the solar limb. Solar south is up. The spatial resolution of IRIS ia about 4 times better than that of AIA. Figure made by the authors from IRIS and AIA data. 
}
\label{spicules}
\end{figure}

The form of dark mottles, particularly in images far from the center of the disk, suggests predominantly radial structures; they appear as {\it spicules} beyond the limb (Figure~\ref{spicules}); they are slender (300-1000 km) plasma jets with speeds of tens of km\,s$^{-1}$ to $>100$ km\,s$^{-1}$, which protrude well up into the corona, reaching heights above 10\,Mm \citep{1957AnAp...20..179M,  1968SoPh....3..367B, 1972ARA&A..10...73B, Tsiropoula2012}; see also bottom panel of Figure 9 and Figure 10 in \citet{2020FrASS...7...74A}). Ordinary spicules have life times of $\sim 10$ min, while of order $10^7$ spicules are active on the Sun at any given time. They collectively carry $\sim\!100$ times the mass flux of the solar wind into the low solar atmosphere \citep{1968SoPh....3..367B, 1972ARA&A..10...73B}).  The mechanisms of their origin and their role in the energy and mass budgets of the outer solar atmosphere and solar wind is therefore a topic of considerable interest. Recent progress on refining their physical properties has largely been driven by O/UV observations from space-based instrumentation: {\sl Hinode}/SOT and IRIS. These have led to the classification of two classes of spicules, type I and type II \citep{2007PASJ...59S.655D} the latter being prevalent in coronal holes and the quiet Sun. Type II spicules are narrower, short-lived (a few tens of s) and faster than type I spicules.

Spicules are much more extended in the He{\sc\,ii} 304\,\AA\ line, where they are usually referred to as {\it macrospicules} \citep{1975ApJ...197L.133B}. In spite of many decades of research, the origin of spicules is still a subject of debate. As disk mottles are clearly associated with the network, a magnetic association is very likely; what is not clear is whether they are a result of reconnection, as suggested a long time ago by \citet{1969AZh....46..328P}, see also \citet{2019Sci...366..890S}, or some other mechanism, such as the leakage of photospheric oscillations expelling plasma along the magnetic field lines, \citep{2004Natur.430..536D}, or even a consequence of ambipolar diffusion \citep{2017Sci...356.1269M, 2019ARA&A..57..189C}.

It turns out that the quiet Sun is not that quiet after all. As the spatial and temporal resolution of our instruments improve, more and more kinds of small-scale transient events are detected (see, e.g., the recent works by \citet{2022A&A...657A.132K, 2022A&A...659A..52S, 2022ApJ...926...42S, 2022A&A...661A.149P}). Such events have been given various names, too many to mention them all, and a possible source of confusion. In the EUV, events with energies as low as $10^{24}$\,erg (e.g. \citet{2022A&A...661A.149P}) have been reported, which puts them inside the range of {\it nanoflares}, proposed by \citet{1988ApJ...330..474P} as the mechanism of coronal heating (see Section 7 in \citet{2020FrASS...7...74A} for a brief discussion). ALMA observations have revealed many such events that are discussed in detail in the review by Nindos et al., included in this special Research Topic collection.

\section{Why ALMA?}\label{ALMA?}

Continuum observations of the quiet Sun at mm-$\lambda$ and submm-$\lambda$ have been performed for decades, serving as a powerful complement to optical O/UV/EUV observations. Indeed, \cite{1973SoPh...28..409L}  compiled quiet Sun measurements from $\lambda$=1-20 mm made during the 1960s and cross-calibrated them against the Moon, resulting in a useful compendium used in models for many years; e.g., the well-known semi-empirical models of \citet{1981ApJS...45..635V}. The primary advantages of observations in the mm/submm-$\lambda$ regime are well known: 1) The sources of opacity are well understood: at mm-$\lambda$ the opacity is dominated by H and He free-free absorption; at submm-$\lambda$ H$^-$ free-free absorption (collisions between electrons and neutral hydrogen) becomes non-negligible and is an additional source of opacity. 2) The radiation at these wavelengths is emitted under conditions of local thermodynamic equilibrium and the source function is therefore Planckian. Furthermore, the Rayleigh-Jeans approximation is valid and the observed intensity (or brightness temperature) is linearly proportional to the kinetic temperature of the emitting plasma for optically thick emission. \citet{2017ApJ...845L..19B} pointed out, however, that the degree of ionization for H and He can depart significantly from conditions expected for ionization equilibrium. As a consequence, while the source function can be in LTE, the opacity may be far from its LTE value. 3) Finally, the radiative transfer of continuum radiation at these wavelengths is straightforward. 

The obvious and persistent disadvantage to observing in the mm/submm-$\lambda$ regime has been that observations have largely been made with single dishes for which the angular resolution was poor (see \citet{1973SoPh...28..409L} and references therein for early measurements as well as  \citet{1984ApJ...281..862L, 1986ApJ...308..448L, 1990ApJ...353L..53L,
1992Natur.358..308L, 1993ApJ...415..364B}. Occasionally, interferometers with limited numbers of antennas have been used to observe the Sun as a means of improving the the angular resolution (e.g., \citet{1981ApJ...244..340H}, White et al. 1996), but the ability to image was absent or relatively poor. Eclipse observations with single dishes or interferometers have also been exploited to garner high resolution information, but it was necessarily in one dimension \citep{ 1983ApJ...264L..25L, 1991ApJ...381..288R, 1992ApJ...400..692B}. In addition to the limitations imposed by poor angular resolution, time-resolved observations were infrequent.

\citet{Gary1996},  \citet{2004A&A...419..747L, Loukitcheva2015}, and \citet{Benz2009} reviewed many pre-ALMA quiet Sun brightness temperature measurements and compared the observational results with the results of various theoretical models. The quiet Sun emission in the wavelength range from $\lambda = 0.85$ mm to $\lambda = 8$ mm was measured by \citet{1993ApJ...415..364B},  \citet{2006A&A...456..697W}, \citet{Brajsa2007a, Brajsa2007b}, and  \citet{2017SoPh..292...22I} and in the wavelength range 0.7 mm to 5 mm various measurements were summarized by \citet{2017SoPh..292...88W}. In the wavelength range 0.7 mm to 5 mm the quiet Sun brightness temperature varies from 5000 K to 8000 K \citep{2017SoPh..292...88W}. As expected, there is an average trend of the brightness temperature increase with wavelength, although some scatter of the results and measurement uncertainties are present.

Interferometric maps of the quiet Sun in the mm-range were first produced by \citet{2006A&A...456..697W} and \cite{2006A&A...456..713L}. They used the 10-element Berkeley-Illinois-Maryland Association Array (BIMA) in its most compact D-configuration, to obtain $\sim10$\arcsec\ resolution. Images from BIMA and the Combined Array for Research in Millimeter-wave Astronomy (CARMA) at 3.5 and 3.3 mm wavelengths are shown in Figure 5 of \citet{2011SoPh..273..309S}. In all cases the chromospheric network, delineated in the TRACE continuum images or the photospheric magnetograms, is the dominant structure in the radio images.

ALMA represents a significant step forward (see \citet{2016SSRv..200....1W}). For the first time it provides the scientific community with high-fidelity imaging of the Sun at sub-arcsecond angular resolution, allowing detailed comparisons with images made in the O/UV/EUV wavelength regimes. In addition, it does so with an imaging cadence as short as 1~s. ALMA therefore offers a fundamentally new tool for studying the quiet Sun chromosphere in a manner that is fully commensurate with the capabilities currently available at other wavelengths. ALMA therefore offers powerful new observations of the quiet solar chromosphere that, particularly when made jointly with O/UV/EUV observations, provide new insights and constraints on our understanding of chromospheric phenomena. Bastian et al, in a review article included in this special Research Topic collection, provide a comprehensive summary of current ALMA capabilities for solar observing as well as capabilities anticipated to be available in the coming years. Technical details regarding solar observing with ALMA are given by  \citet{2017SoPh..292...87S} and \citet{2017SoPh..292...88W}.

\section{Full-disk ALMA observations}\label{Section:FD}
ALMA provides both high resolution interferometric imaging and lower resolution full-disk imaging. The latter does not only serve to determine the background brightness for interferometric images, but also give a global view of the Sun, not available in the high resolution images due to their small field of view. Thus full-disk images give information on all kinds of solar phenomena not yet imaged in interferometric mode; moreover, they provide accurate measurements on the variation of the brightness temperature from the center of the disk to the limb which, in turn, is important for solar atmospheric models.

\subsection{Morphology and comparison with other wavelength ranges}\label{morph}

Several full-disk images were obtained during the ALMA solar commissioning observations in December 2015 at 239 and 100\,MHz. \cite{2017A&A...605A..78A} compared  images obtained on December 17 in both bands with \ha\ images from the GONG network, 1600\,\AA\ images from AIA/SDO and magnetograms from HMI/SDO and noted that plage regions were the most prominent feature on the disk. A large sunspot was clearly visible at 239 and barely discernible at 100 GHz. Prominences were well visible beyond the limb, but large-scale neutral lines rather than filaments were seen on the disk, as darker-than-average features. \begin{figure}[h]
\begin{center}
\includegraphics[width=.6\textwidth]{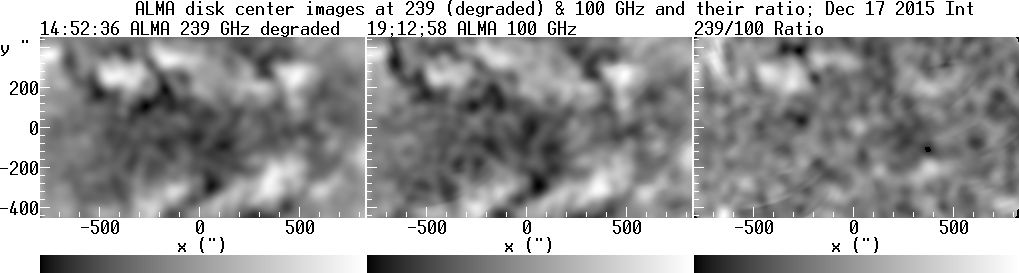}
\end{center}
\caption{
A region near the center of the disk observed with ALMA in full-disk mode on December 18, 2015: Degraded 239\,GHz image, 100\,GHz image, and their ratio. Values range from 6000 to 7380\,K, 6860 to 8380\,K and 0.83 to 0.93, respectively. From \cite{2017A&A...605A..78A}, reproduced with permission \copyright\ ESO.}
\label{ratio}
\end{figure}

\begin{figure}[h]
\begin{center}
\includegraphics[width=.9\textwidth]{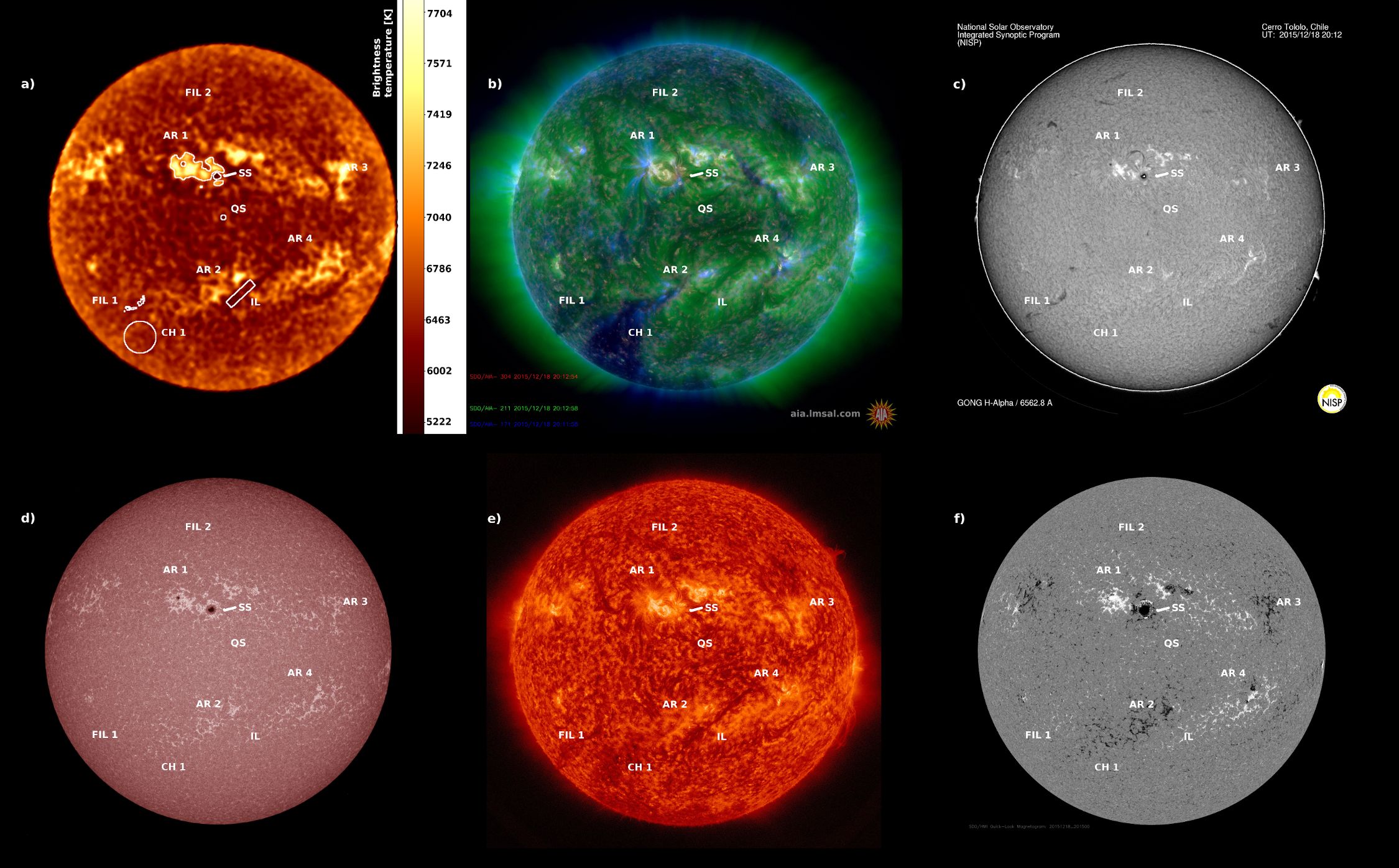}
\end{center}
\caption{Images of the Sun from different instruments taken on December 18, 2015. {\bf(Top row)} a): ALMA intensity map at 248 GHz ($\lambda = 1.21$ mm, 20h 12m 21s). The brightness temperature in K is given on the intensity bar on the right. 
 b): SDO composite image from AIA 30.4 nm, AIA 21.1 nm, AIA 17.1 nm instruments (20h 12m 58s UT). c):  \ha\ filtergram from Cerro Tololo Observatory, NISP (20h 12m UT). {\bf(Bottom row)} d): SDO AIA 170.0 nm filtergram. e): SDO AIA 30.4 nm filtergram.  f): SDO HMI magnetic field. 
Several regions of interest are encircled by white lines: AR indicates the position of active regions, FIL shows the position of filaments, SS shows the position of the sunspot, QS shows the position of the central quiet Sun region, IL indicates the position of magnetic inversion line, while CH is the position of the coronal hole. Adapted from \citet{2018A&A...613A..17B}. Reproduced with permission \copyright\ ESO.
}
\label{Roman_1}
\end{figure}

\begin{figure}[h]
\begin{center}
\includegraphics[width=14cm]{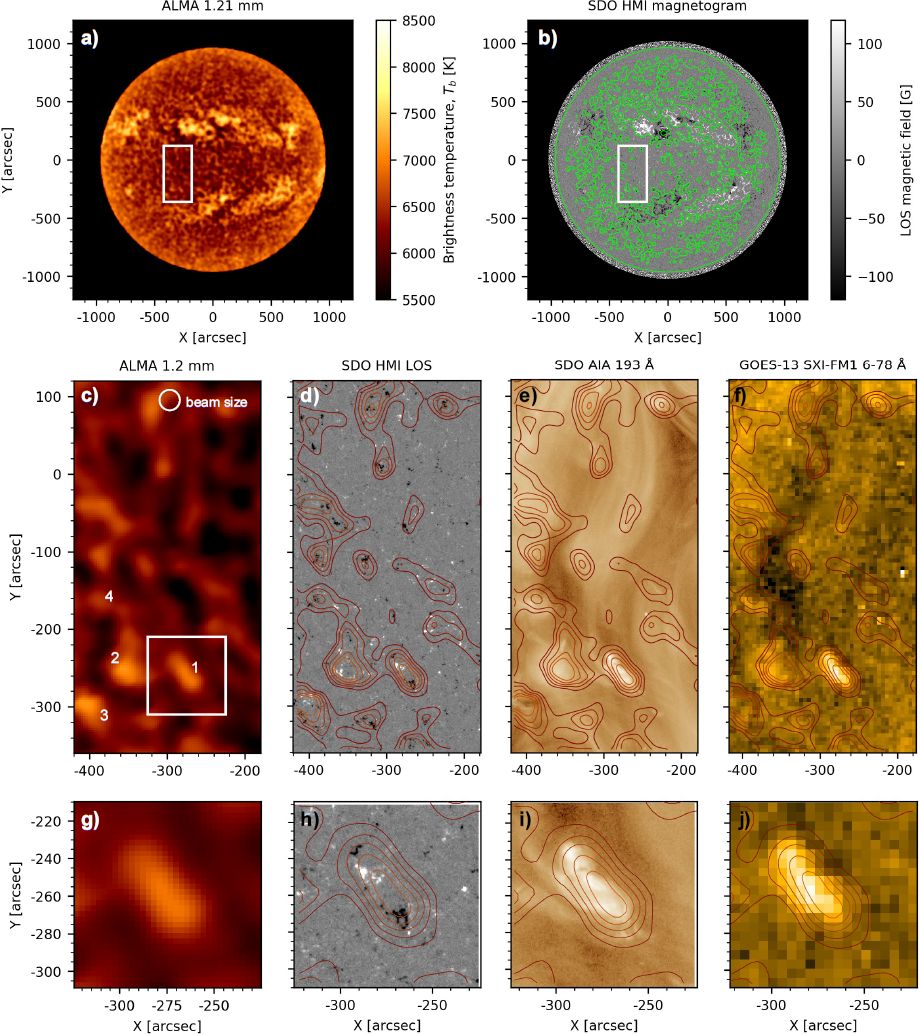}
\end{center}
\caption{
{\bf Different views of coronal bright points: 1.21 mm (ALMA), magnetic field, EUV, soft X-rays.}
	(a): Single-dish ALMA 248 GHz (1.21 mm, Band 6),    
	(b): SDO/HMI magnetogram,  
	with white box outlining the quiet Sun area shown in panels in the middle row. 
	Both images were taken on 18 December 2015 at 20:12 UT. Magnetogram intensity was clipped at $\pm$120 Gauss 
	 and ALMA contours of 6500 K were overlaid in green color.
	(c):  the ALMA image of the quiet Sun shown enlarged and compared with 
	(d): SDO/HMI magnetogram, 
	(e): SDO/AIA 193~\AA~filtergram, and
	(f): GOES-13 SXI tin (Sn) filter image, with ALMA contours overlaid  (levels at 6300, 6400, 6500, 6600 and 6700 K).
	(g-j): enlargements of a single ALMA bright feature, marked  in the white rectangle in panel (c), 
		at the same wavelengths as in the middle row. 
		The beam size of the  single-dish ALMA measurements, 26", is given in the panel (c). 
Reproduced from \citet{2021A&A...651A...6B} with permission \copyright\ ESO.
}
\label{Roman_2}
\end{figure}

They also noted that the chromospheric network is well visible at 239 GHz and very similar to the AIA 1600 and 304\,\AA\ images, with a slightly better correlation with 1600\,\AA. Moreover, they computed the 239/100 GHz intensity ratio, after smoothing the 239\,GHz images to the 100\,GHz resolution (Figure~\ref{ratio}) and reported a range of values from from 0.83 to 0.93 and a spectral index in the range of $-0.21$ to $-0.083$. The ratio was higher for the active region plage in the upper part of the figure, which had a flat spectrum, but not for the faculae in the lower part of the figure, whereas the dark lanes between the faculae were indistinguishable in the ratio image. \cite{2020A&A...640A..57A} made similar comparisons with Band 7 (0.85\,mm) ALMA images (see their Figures 1 and 2).

\citet{2018A&A...613A..17B} analyzed a full-disc solar ALMA image at 1.21 mm obtained on December 18, 2015, during a CSV-EOC  (Commissioning and Science Verification - Extension of Capabilities) campaign. The ALMA image was calibrated and compared with full-disc solar images from the same day in H$\alpha$ line, in He I 1083 nm line core, and with various SDO images (AIA at 170 nm, 30.4 nm, 21.1 nm, 19.3 nm, and 17.1 nm and HMI magnetogram). The brightness temperatures of various structures were determined by averaging over corresponding regions of interest in the calibrated ALMA image. Positions of the quiet Sun, active regions, prominences on the disc, magnetic inversion lines, coronal holes and coronal bright points were identified in the ALMA  image (Figure~\ref{Roman_1}). At the wavelength of 1.21 mm, active regions appear as bright areas (but sunspots are dark), while prominences on the disc and coronal holes are not discernible from the quiet Sun background, despite having slightly less intensity than surrounding quiet Sun regions. Magnetic inversion lines appear as large, elongated dark structures and coronal bright points correspond to ALMA bright points. The identification of coronal bright points in the ALMA image was one of the most important new result of that study. The great majority of all coronal bright points from the EUV image correspond to the He I 1083 nm dark points (75 \%) and to the ALMA 1.21 mm bright points (82 \%). Moreover, all ALMA 1.21 mm bright points show a well-defined relationship with magnetic structures (100 \% correspondence), mostly with small-scale bipolar magnetic regions and in some cases with small unipolar magnetic regions.

\citet{2021A&A...651A...6B} investigated small-scale ALMA bright features in the quiet Sun region using a full-disk solar image produced with single-dish ALMA measurements (1.21 mm, 248 GHz) performed on December 18, 2015. The selected quiet Sun region was compared with the EUV and soft X-ray images and with the magnetograms (Figure~\ref{Roman_2}). 
We note that with the GOES-13 SXI (Solar X-ray Imager)-FM1 filter soft X-rays are detected, which are caused by the bremsstrahlung of the hot plasma from the chromosphere and corona. In the quiet Sun region, enhanced emission seen in the ALMA was almost always associated with a strong line-of-sight magnetic field. Four coronal bright points were identified and one typical case was studied in detail (Figure~\ref{Roman_2}). Other small-scale ALMA bright features were most likely associated with magnetic network elements and plages.

The question of visibility of features in \ha, Ly-$\alpha$ and ALMA has been addressed by \citeauthor{2017A&A...598A..89R} (\citeyear{2017A&A...598A..89R}; see also \citet{2017A&A...597A.138R, 2019A&A...632A..96R}, under the working hypothesis that the observed ubiquitous long \ha\ fibrils are contrails from propagating heating events. Application of a recipe that assumes momentary Saha-Boltzmann extinction during their hot onset to millimeter wavelengths, suggests that ALMA should observe \ha-like fibril canopies, not acoustic shocks underneath. Therefore, anything opaque in H$\alpha$ will be at least similarly opaque at mm wavelengths. \citet{2017A&A...598A..89R} concluded that mm contrail fibrils would be yet more opaque than the H$\alpha$ ones, constituting yet denser canopies. In short, \citet{2017A&A...598A..89R} predicted that the 
ALMA quiet-Sun is mostly chromospheric canopies, not appearing the same as in \ha, but opaque as in \ha.

A good dark-dark correspondence (sunspot and small filaments) and bright-bright correspondence (plages) between the two images was indeed found by \citet{2021A&A...651A...6B}. However, the relationship between H$\alpha$ core and high-resolution ALMA images in the millimeter wavelength range is more complicated, since these authors found examples of both types of behavior, correlation and anticorrelation. Some small-scale, bright ALMA features have bright H$\alpha$ counterparts, and some have dark H$\alpha$ counterparts (including fibrils). Moreover, examples of a bright-dark and a dark-bright correlation between the H$\alpha$ wing sum and ALMA image were reported. We can thus conclude that, while some of the predictions of \citet{2017A&A...598A..89R} have been qualitatively confirmed using ALMA observations, there are notable exceptions.

\subsection{Center-to-limb variation and atmospheric models}\label{Section:CLV}
\begin{figure}%[h]
\begin{center}
\includegraphics[height=8.5cm]{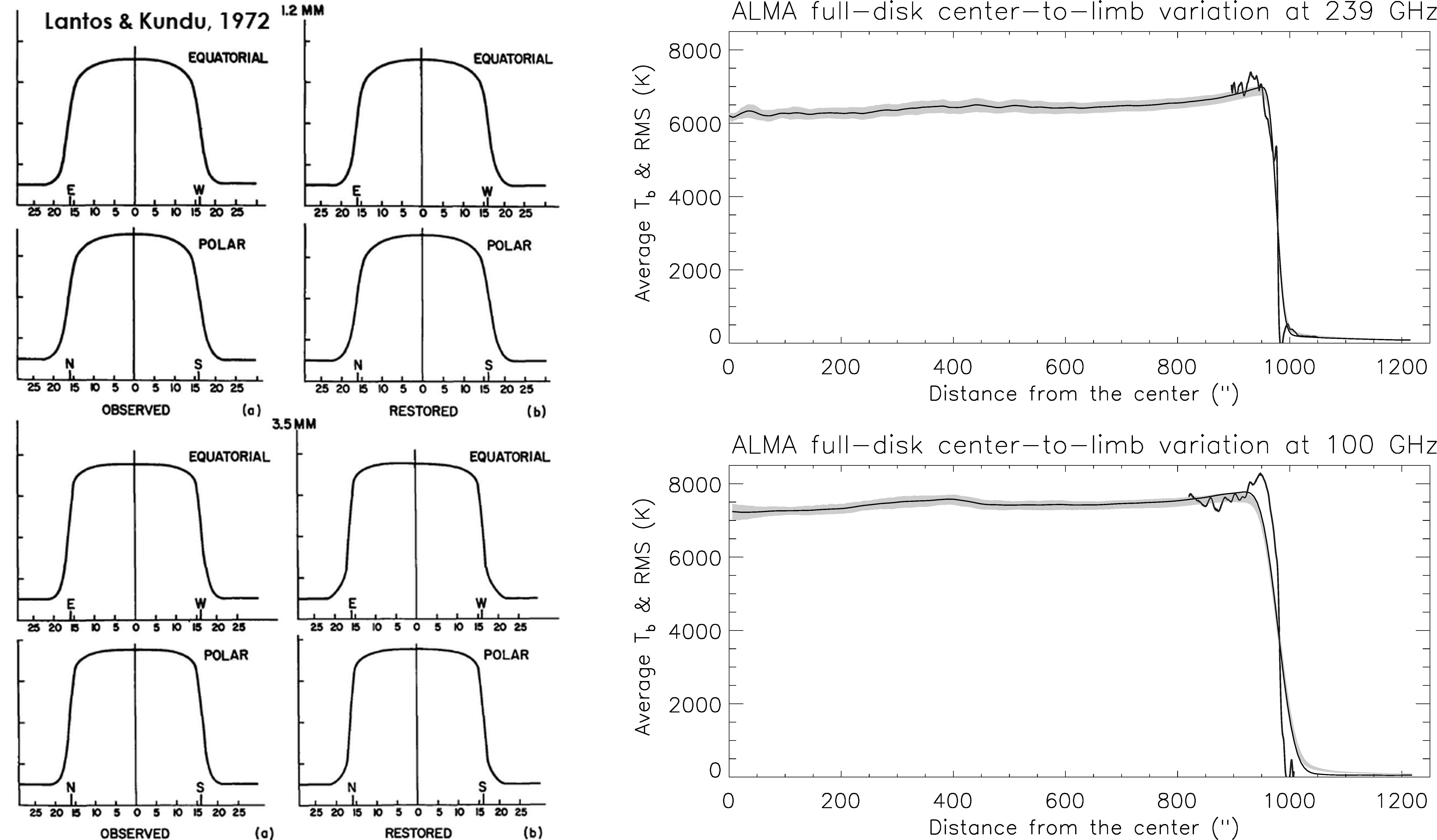}
\end{center}
\caption{Old and new Center-to-limb variation (CLV) curves) {\bf Left}: CLV curves at 1.2 and 3\,mm (250 and 100\,GHz respectively) from \cite{1972A&A....21..119L}.  Reproduced with permission \copyright\ ESO. {\bf Right}: CLV curves for 239 and 100 GHz form ALMA full-disk images from \cite{2017A&A...605A..78A}. The gray band shows the measured values $\pm$ the corresponding rms. The superposed blue curves near the limb are from high resolution ALMA images. Curves in black show the CLV after correction for diffuse light. Reproduced with permission \copyright\ ESO.
}
\label{CLV}
\end{figure}

The first imaging observations of the Sun in the mm range were made in the 70's, with the 36-ft radio telescope of the National Radio Astronomy Observatory in Tucson, Arizona \citep{1970A&A.....5..102B, 1970SoPh...13..348K, 1971SoPh...21..130K, 1972SoPh...25..108K, 1972SoPh...24..133K}). This instrument provided resolution of 3.5\arcmin, 1.2\arcmin\ and 1.2\arcmin\  at 9\,mm, 3.5\,mm and 1.2\,mm respectively. Analyzing the center-to-limb variation (CLV) of the brightness temperature and taking into account the effect of the instrumental beam, \citet{1972A&A....21..119L} found limb darkening both at 1.2 and 3\,mm (Figure~\ref{CLV}, left). As solar atmospheric models predict limb brightening at these wavelengths, \citet{1972A&A....21..119L} had to invoke absorbing structures (spicules) in order to match the models with the observations.

ALMA full-disk images are much more precise than the old 36-ft images and the first analysis of commissioning observations by \cite{2017A&A...605A..78A} showed clear limb brightening both in Band 3 (100\,MHz, 3\,mm) and in Band 6 (250\,GHz, 1.26\,mm), available at that time (right panel of Figure~\ref{CLV}). Moreover, \cite{2017A&A...605A..78A} combined the measurements of brightness temperature, $T_b$, as a function of the cosine of the heliocentric distance, $\mu$, from both frequency bands to a single curve $T_b(\mu_{ref})$, by reducing all values to a common reference frequency of 100\,GHz; they also extended their frequency range by including measurements by \cite{1993ApJ...415..364B} at 350\,GHz and found a linear relation between $T_b$ and $\mu_{100}$. Subsequently, they inverted the equation of transfer,  

\begin{equation}
T_b(\mu)=\int_0^\infty T_e(\tau)\, e^{-\tau/\mu}\,d\tau/\mu \label{transf}
\end{equation}

\noindent where $\tau$ is the optical depth, to obtain the electron temperature, $T_e$ as a function of $\tau_{ref}$.

The inversion was done by noting that a linear (or polynomial) $T_e(\log\tau_{100})$ relation leads also to a linear (or polynomial) $T_b(\log\mu_{100})$ relation, thus the coefficients of the former can be computed from the coefficients of the latter, determined from a fit to the observed data. For example, a 3rd degree polynomial expression for the electron temperature:

\begin{equation}
T_e(\tau)=a_0+a_1\ln\tau+a_2\ln^2\tau+a_3\ln^3\tau \label{logform},
\end{equation}

\noindent gives a brightness temperature of:

\begin{eqnarray}\label{A12}
T_b(\mu)&=&a_0 + a_1\,C_1 + a_2\,C_2 + a_3\,C_3 \\ \nonumber
&+& (a_1 + 2a_2C_1 + 3a_3C_2)\ln\mu \\ \nonumber
&+& (a_2 +3 a_3\,C_1)\ln^2\mu\\ \nonumber
&+& a_3\ln^3\mu, 
\end{eqnarray}

\noindent with

\begin{eqnarray}
C_1&=&-0.577216,\\
C_2&=&~~~1.978112, \hspace{0.3cm}\mbox{ and}\\
C_3&=&-5.444874
\end{eqnarray}

\noindent (see Appendix 1 in \citet{2022A&A...661L...4A}.

A similar approach was employed by \citet{2020A&A...640A..57A} (see their Figure 4 for a comparison of the observed CLV with the model); these authors cross-calibrated band 6 with respect to band 3 using the CLV curves, something that was not necessary for the data used by \citet{2017A&A...605A..78A}.

We note that the parameters derived from the fit can be used for the computation of the brightness temperature spectrum at the center of the solar disc. Such a computation was done by \citet{2020A&A...640A..57A}, see their Figure 3, and the derived spectrum was compared with older measurements compiled by \citet{Loukitcheva2004}, which is a highly inhomogeneous data set and has a lot of scatter. Several points were near the ALMA
spectrum, but most of them were below.

\begin{figure}[h]
\begin{center}
\includegraphics[height=8.cm]{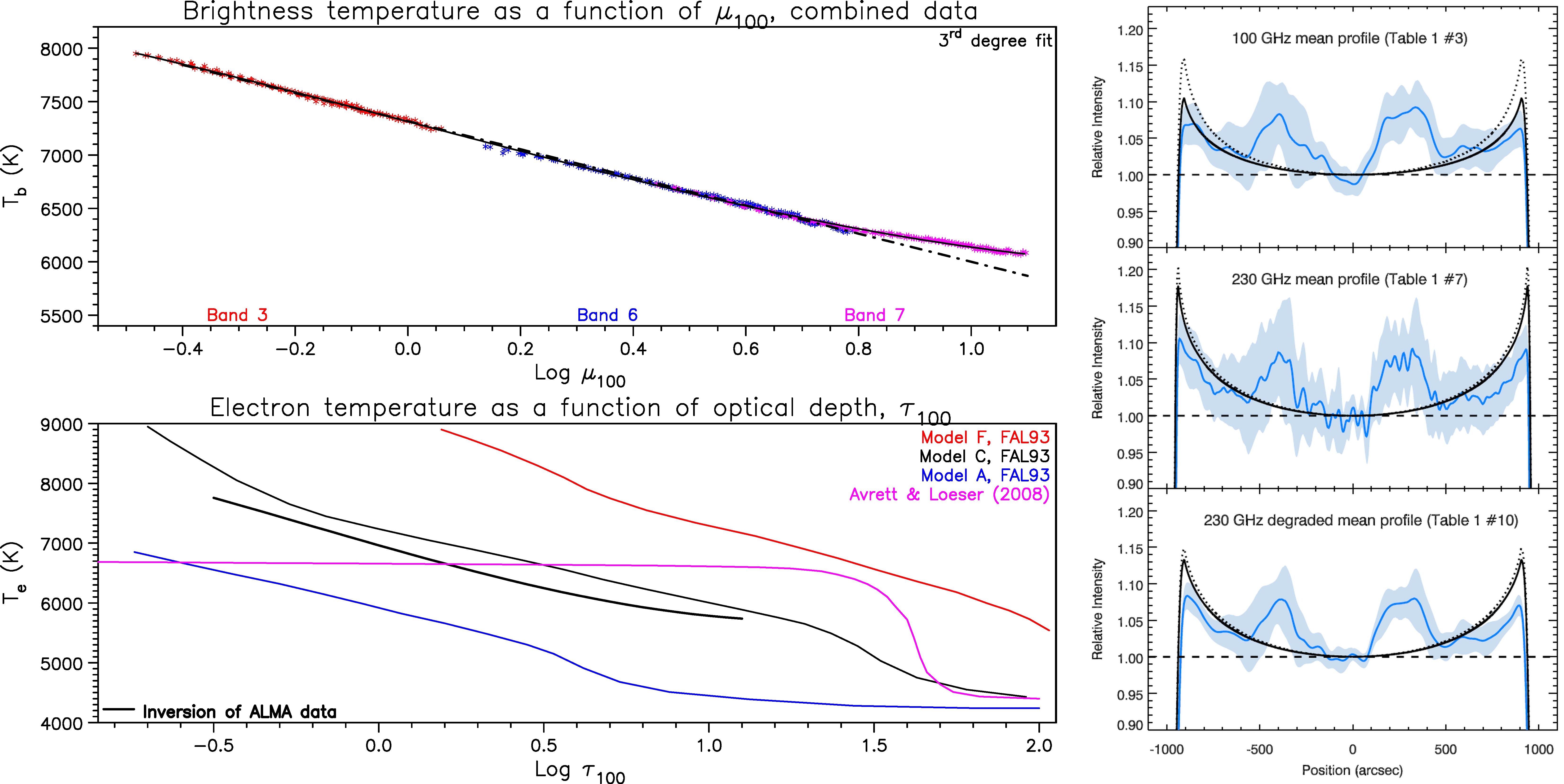}
\end{center}
\caption{Atmospheric modeling from ALMA CLV data. {\bf Left, top}: Combined CLV curve, $T_e(\log\tau_{100})$, from ALMA Band 3, Band 6 and Band 7 observations obtained in January 2020, adapted from \citet{2022A&A...661L...4A}. The solid line shows a third degree fit and the dash-dotted line shows a linear fit up to $\log\mu_{100} = 0.8$. Reproduced with permission \copyright\ ESO. {\bf Left, bottom}: The electron temperature as a function of optical depth (thick black line) obtained from the inversion of the CLV curve. Also shown are the predictions of atmospheric models discussed in the text. Adapted from \citet{2022A&A...661L...4A}. {\bf Right:} Comparison between observed CLV (blue curves) and the predictions of models C (dotted lines) and SSC (continuous black lines). The dashed line represents the quiet Sun. From \citet{2019ApJ...871...45S} \copyright\ AAS, reproduced with permission.
}
\label{invert}
\end{figure}

In a recent work \citet{2022A&A...661L...4A} used ALMA Band 7 (347\,GHz, 0.86\,mm) data instead of the measurements by \citet{1993ApJ...415..364B}; this, more homogeneous data set, revealed a flattening of the $T_b(\log\mu_{100})$ curve at large $\mu$, with a 3rd degree polynomial fitting the data better. The results are shown in Figure~\ref{invert} (left). The left top panel gives the CLV curve derived from the combination of all three available ALMA frequencies and its fit with a 3rd degree curve. The thick black curve in the left bottom panel shows the result of the inversion, which is compared to the predictions of models A (internetork), C (average quiet Sun) and F (network) of \citet{1993ApJ...406..319F}, as well to model C7 of \citet{2008ApJS..175..229A}. The physical parameters of these models were used to compute $\tau_{100}$ as a function of the height, $z$, and, though the $T_e(z)$ relation, the optical depth as a function of $T_e$. 

The observational result is very close to the prediction of model C of the  \citet{1993ApJ...406..319F}, as already noted by \citet{2017A&A...605A..78A}, and very different from that of the \citet{2008ApJS..175..229A} model, which predicts too small a temperature  variation in the chromosphere. The flattening at high optical depths is probably an indication that we approach the region of temperature minimum. As the observations are very close to the model, there is no need to invoke spicule absorption.

In a different approach, \citet{2019ApJ...871...45S} compared their CLV measurements of the commissioning observations with model C of \citet{1993ApJ...406..319F} and with their own SSC model \citep{2005A&A...433..365S}. As shown in Figure~\ref{invert} (right), they found that the SSC model gave a better fit to the observations. The same authors reported average polar brightening of 10.5\% and 17.8\% at 100 and 230\,GHz, respectively. Finally, \citet{2019SoPh..294..163S} fitted the CLV, again from the commissioning data, with a quadratic curve in $\mu$ and used their results for image corrections, without performing any comparison with model predictions.

\subsection{Formation height and the solar radius}\label{Sect:height}
In order to pass from the observationally determined $T_e(\tau)$ to the variation of temperature as a function of height, $T_e(z)$, we need a relation between $\tau$ and $z$. This implicates the opacity and hence the density as a function of height, for which we have no direct observational information; the situation is further complicated by the fact that temperature variations change the degree of ionization and hence the electron density, in a way that is complicated due to departures from local thermodynamic equilibrium (LTE). Thus, no attempt to derive $T_e(z)$ from mm-$\lambda$ $T_e(\tau)$ has been made yet.

Qualitative information about the formation height can be provided through the comparison of the morphology of ALMA images with images in other spectral regions. Already \citet{2017A&A...605A..78A} noted the similarity of chromospheric network structures in ALMA and AIA 1600\,\AA\ and 304\,\AA\ images, putting their formation height between the two; this is consistent with the fact that AIA 304\,\AA\ channel is dominated by the He{\sc\,ii} lines which form around $T_e=50\,000$\,K \citep{2010A&A...521A..21O}, much above the mm-$\lambda$ brightness temperature. We note that, in the cm-$\lambda$ range, where radiation forms higher, \citet{Bogod2015} found a better correlation of RATAN-600 data with 304\,\AA\ AIA images.

Indications about the formation height can be obtained from the association of the observed disk center brightness temperature to the electron temperature predicted by models. For model C of \citet{1993ApJ...406..319F}, this gives heights of 1.09, 1.25 and 1.82\,Mm above the $\tau_{5000}=1$ level, which is the level where the optical depth at 5000\,\AA\ is unity, for radiation at 0.85, 1.26 and 3\,mm, respectively. 

The solar radius is another indicator of the formation height; however the radius reflects the maximum height of formation rather than the average, the presence of spicules complicate the measurement, while it cannot be accurately determined from the relatively low resolution full-disk ALMA images. \citet{2020A&A...640A..57A} compiled various height measurements in their Table 4; they gave heights of $2.4\pm1.7$\,Mm and $4.2\pm2.5$\,Mm above the $\tau_{5000}=1$ level,  for the limb at 1.26 and 3\,mm respectively. Both are between the measured heights of the 1600\,\AA\ and the 304\,\AA\ emissions, but below the heights measured in eclipse observations. \citet{2019ApJ...871...45S} obtained average values of the solar radius of $965.9\pm3.2$\arcsec\ at 3\,mm and $961.6\pm2.1$\arcsec\ at 1.26\,mm, which correspond to heights above the $\tau_{5000}=1$ level of $4.6\pm2.3$\,Mm and $1.5\pm1.5$\,Mm respectively and are within the error margins of \citet{2020A&A...640A..57A}. More recently, \citet{2022MNRAS.511..877M} gave values of $968\pm3$\arcsec\ for the solar radius at 3\,mm  and $963\pm2$\arcsec\ at 1.26\,mm, corresponding to heights of $6.1\pm2.1$\,Mm and $2.5\pm1.4$\,Mm respectively, above the $\tau_{5000}=1$ level.

\section{High resolution ALMA observations}\label{Section:HR}
\begin{figure}[h]
\begin{center}
\includegraphics[height=4cm]{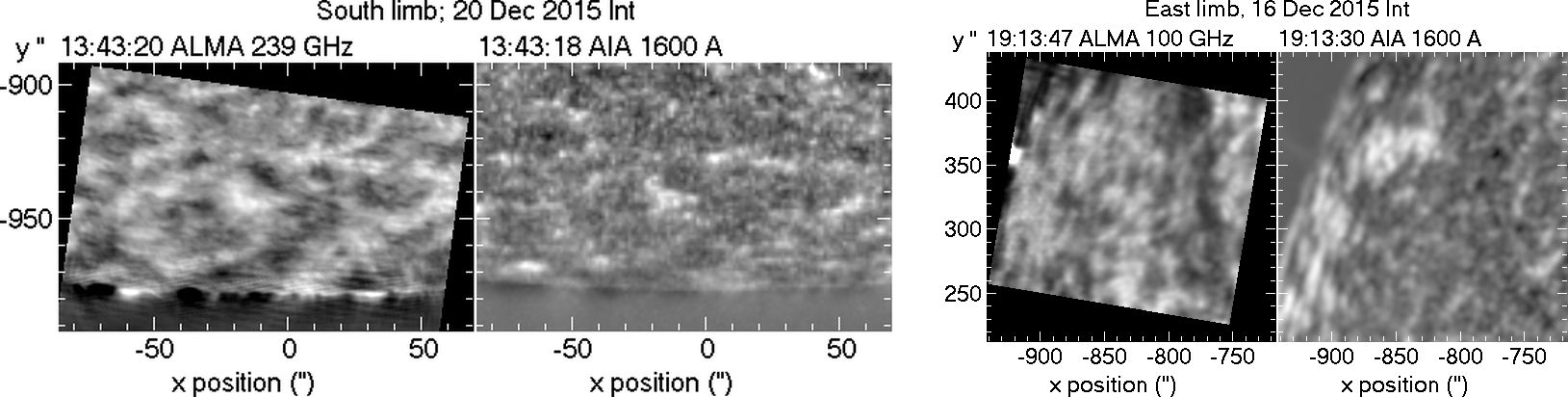}
\end{center}
\caption{ALMA images near the S and E limbs obtained during the commissioning period, together with the corresponding AIA 1600\,\AA\ images. A plage region is visible near the E limb. All images have been corrected for CLV, by subtracting 85\% of the azimuthally averaged intensity; AIA images were smoothed to the ALMA resolution of 1.76 by 1.01\arcsec\ at 239\,GHz and 5.45 by 4.00\arcsec\  at 100\,GHz. Solar north is up. Figure made by the authors from data set ADS/JAO.ALMA.2011.0.00020.SV.}
\label{CommQS}
\end{figure}

Several ALMA observing sessions focused on quiet Sun regions both on the disk and at the limb. The small field of view of ALMA has not allowed a global view of the Sun at high resolution, while, using mosaicing techniques, two larger regions near the limb were imaged during the commissioning period with fields of view of 60\arcsec\ by 60\arcsec\ at 1.26\,mm and 190\arcsec\ by 180\arcsec\ at 3\,mm (Figure~\ref{CommQS}). Most of the published results are at 3\,mm, where atmospheric conditions are more favorable than at 1.26\,mm, whereas no high resolution observations in Band 7 (0.85\.mm), recently made available for solar observing, have been presented yet.

\subsection{Structures on the solar disk and the chromospheric network}\label{Section:disk}

The most complete set, covering 7 regions from the center of the disk to the limb at 3\,mm with a resolution of 2.4\arcsec\ by \begin{figure}[!h]
\begin{center}
\includegraphics[width=.9\textwidth]{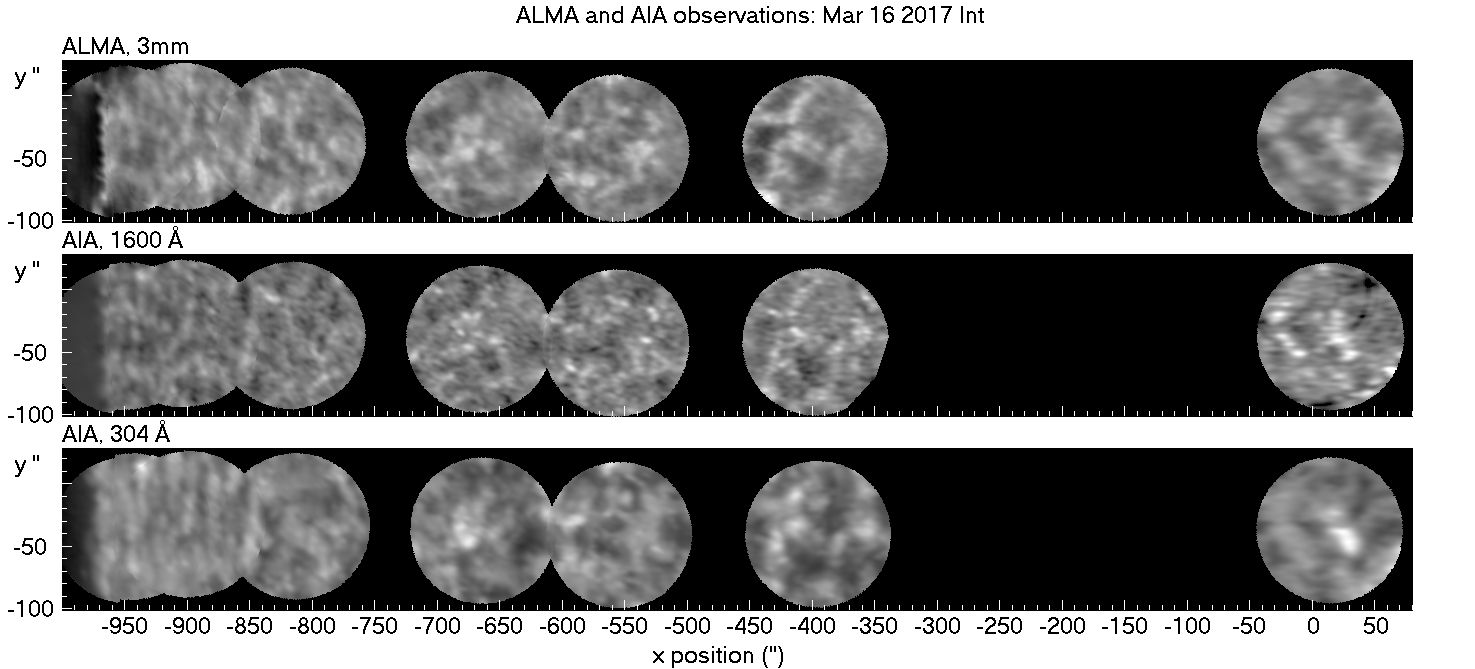}
\end{center}
\caption{ALMA and AIA images of the quiet Sun. {\bf Top row:} Composite of seven 3\,mm ALMA images from the limb (at left) to the center of the disc. {\bf Second and third rows:} Composites of AIA 1600 and 304\,\AA\ images for the same time intervals and FOVs as the ALMA images, convolved with the ALMA beam. The photospheric radius is 964.8\arcsec. From \citet{2018A&A...619L...6N}, reproduced with permission \copyright\ ESO.
}
\label{targets}
\end{figure}
4.5\arcsec, is that of \cite{2018A&A...619L...6N}; these authors extended the ALMA field of view to a diameter of 120\arcsec, twice that of the nominal and were thus able to make better comparisons with simultaneous AIA 1600\,\AA\ and 304\,\AA\ images, smoothed to the ALMA resolution (Figure~\ref{targets}). They thus confirmed the association of the bright mm-$\lambda$ structures to the chromospheric network, discussed in Section~\ref{Section:FD} and also visible in Figure~\ref{CommQS}. In a subsequent work, \cite{2021A&A...652A..92N} presented observations of a very quiet region in both bands 3 and 6, obtained on the same day, with a resolution of 2.7\arcsec\ by 1.7\arcsec\ for band 3 and 1.6\arcsec\ by 0.7\arcsec\ for band 6 (Figure~\ref{ALMAnetwork}). In addition to the similarity with AIA 1600\,\AA\ and 304\,\AA\ images, Figure~\ref{ALMAnetwork} reveals a similarity with negative, broadband, \ha\ images from the GONG network. A similar association of ALMA and \ha\ features was reported by \cite{2020A&A...640A..57A} from full-disk 347\,MHz (0.85\,mm) images.  \cite{2021A&A...652A..92N} attributed  this similarity in terms of absorption in \ha\ by dark mottles (spicules), located above network elements, implying that the mottles are transparent at 3\,mm and that ALMA images show the network points below them. An alternative interpretation, in terms of \citeauthor{2017A&A...598A..89R}'s (\citeyear{2017A&A...598A..89R}) models, is that ALMA sees a part of opaque fibrils. Observationally, it is hard to distinguish between the two; opaque fibrils should appear more elongated than the hitherto images show (see also Figure~\ref{moreALMA2} below), but this could be masked by the ALMA beam (resolution). 

\begin{figure}[h]
\begin{center}
\includegraphics[width=\textwidth]{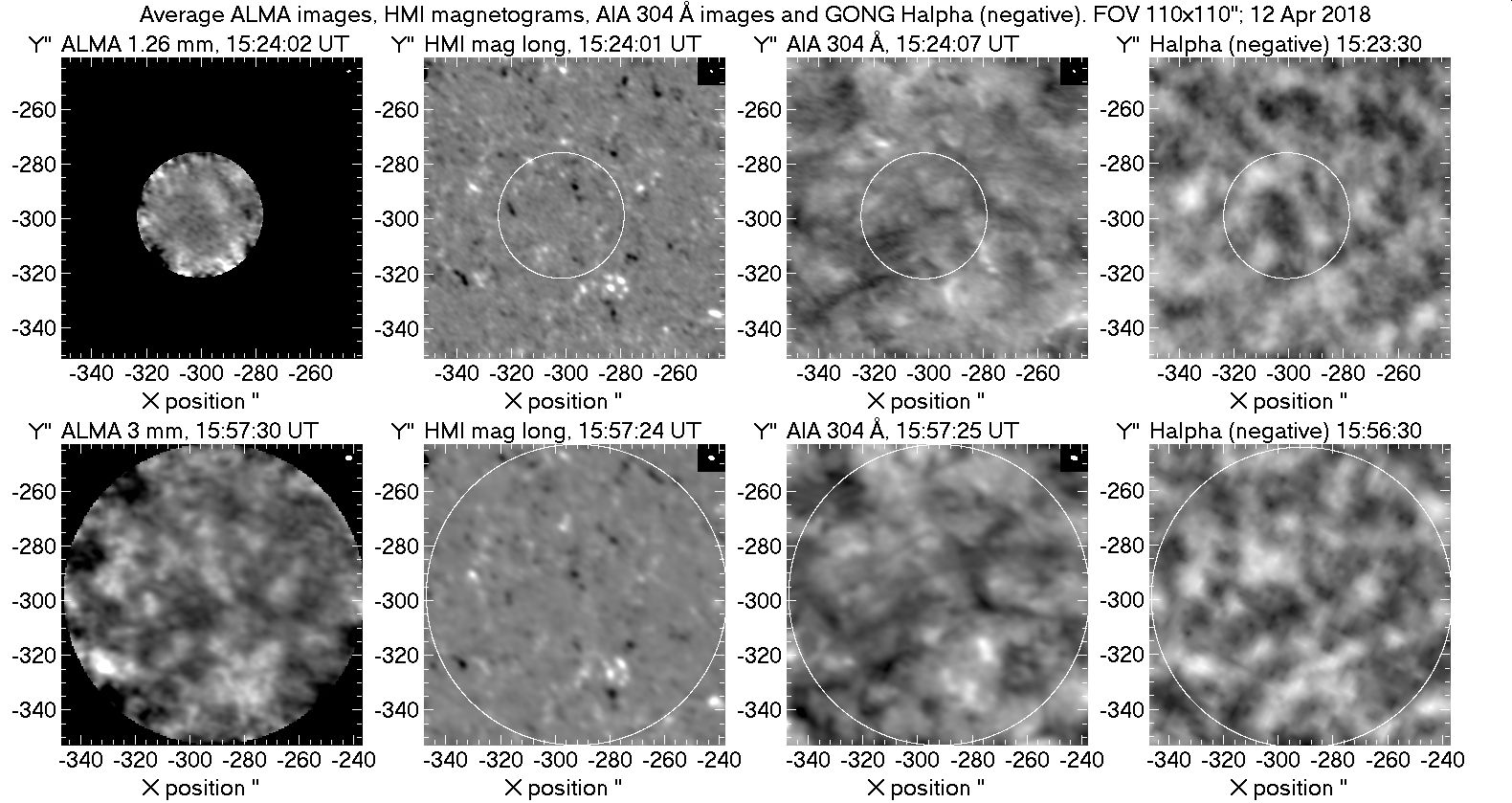}
\end{center}
\caption{Average ALMA images of the network in Band 6 (top-left) and in Band 3 (bottom-left), together with the corresponding HMI magnetograms (saturated at $\pm50$\,G), as well as AIA 304\,\AA and \ha\ negative images from GONG. The insert at the top right corner of the images shows the ALMA resolution. The circles in the non-ALMA images mark the ALMA field of view. All images are oriented with the celestial north pointing up (from \citet{2021A&A...652A..92N}, reproduced with permission \copyright\ ESO).
}
\label{ALMAnetwork}
\end{figure}

\cite{2020A&A...640A..57A} used the observations presented in the previous paragraph, as well as commissioning data (see their Table 6), to measure the CLV of the brightness temperature  for the network and cell interior (internetwork); from that, they computed the electron temperature as a function of the optical depth for these components, as described in Section~\ref{Section:CLV}. As in the case of the average quiet Sun, they found linear relations between $T_e$ and $\log\tau_{100}$. Two segregation schemes were employed, one attributing equal number of pixels in network and cell interior and another, closer to that used in classic models. The latter gave a network/cell temperature difference of $\sim1550$\,K (ratio of $\sim1.24$), considerably less than $\sim3280$\,K (ratio of $\sim1.55$) predicted from models F and A models of \citet{1993ApJ...406..319F}.

\begin{figure}[h]
\begin{center}
\includegraphics[width=\textwidth]{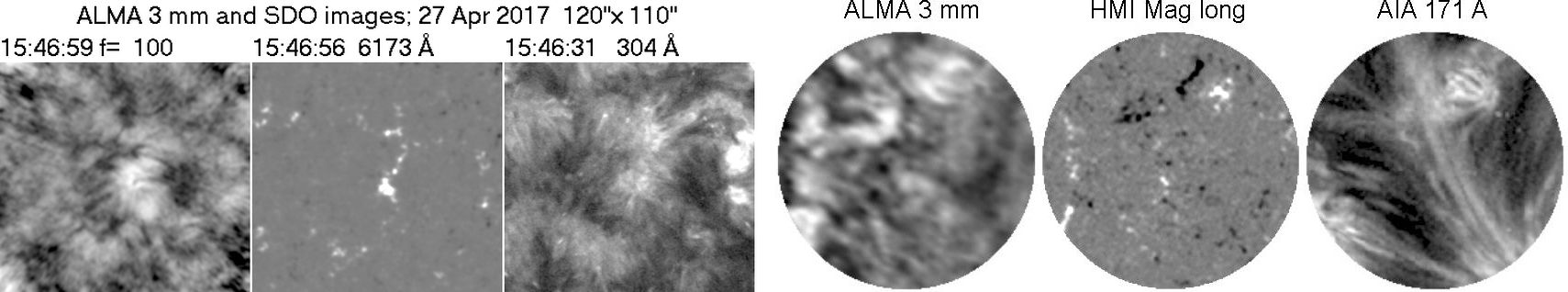}
\end{center}
\caption{ALMA images at 3\,mm. {\bf Left}: A low intensity region at 3\,mm from data used by \citet{2019ApJ...877L..26L}, together with an HMI magnetogram saturated at $\pm200$\,G and a 304\,\AA\ AIA image (prepared by the authors using data from ALMA project 2016.1.00202.S  and from SDO). {\bf Right}: A quiet region at 3\,mm together with an HMI magnetogram and a 171\,\AA\ AIA image; the diameter of the field of view is $\sim67$\arcsec\ (adapted from  \citet{2020A&A...635A..71W}).
}
\label{moreALMA}
\end{figure}

Brightness temperature measurements from interferometric ALMA images at 3\,mm were also published by  \citet{2019ApJ...877L..26L} and \citet{2020A&A...635A..71W}. \citet{2019ApJ...877L..26L} observed a region 200\arcsec\ SW of the disk center with 1.6\arcsec\ resolution and reported an extended low brightness area which they called ``chromospheric ALMA hole'' (Figure~\ref{moreALMA}, left). As a matter of fact, this was the interior of a large supergranule with an average brightness (deduced from the histograms of their Figure 2a) of 6440\,K, considerably below the average cell interior value of 7060\,K used by \citet{2020A&A...640A..57A}, whereas the average brightness of what they called ``bright network'' was 7850\,K, compared to 7690\,K of \citet{2020A&A...640A..57A}. Another region near the center of the disk was observed with a resolution of 1.4\arcsec\ by 2.1\arcsec\ by \citet{2020A&A...635A..71W} (Figure~\ref{moreALMA}, right), who reported average $T_b$ values of 7228 and 7588\,K, for cell and network respectively, while their radiative MHD model predicted average values of 6688 and 7977\,K  for the cell and network respectively.

\begin{figure}[h]
\begin{center}
\includegraphics[width=\textwidth]{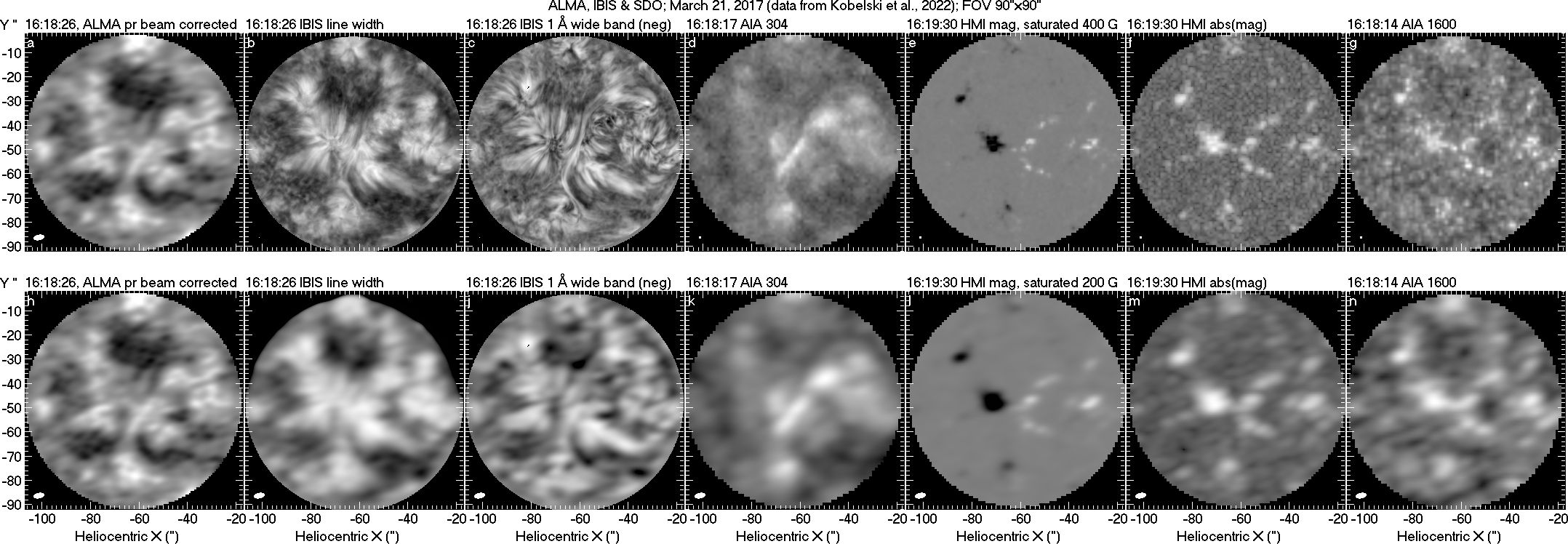}
\end{center}
\caption{ALMA images at 3\,mm. {\bf Top row:} ALMA image together with IBIS \ha\ images and SDO images; the insert at the lower left corner shows the instrumental resolution (beam). {\bf Bottom row:} Same as in middle row, with the images smoothed to the ALMA resolution. Produced by the authors using data from \cite{2022ApJS..261...15K} and SDO.
}
\label{moreALMA2}
\end{figure}

In a recent work, \cite{2022ApJS..261...15K} analyzed 3\,mm ALMA data of a bipolar region of enhanced network, in conjunction with \ha\ images obtained with the Interferometric BIdimensional Spectrometer (IBIS), mounted on the Dunn Solar Telescope.  They reported a strong association between ALMA emission and \ha\ line width (see also the article by Tarr et al., in this collection), confirming a similar finding by \cite{2019ApJ...881...99M} for a plage region. From the publicly available data set of \cite{2022ApJS..261...15K} and SDO data we compiled a set of images, displayed in Figure~\ref{moreALMA2}. We note that the \ha\ linewidth image (panels b and i) is very similar to the negative broadband (1\,\AA) \ha\ image (panels c and j); this is not unexpected, as dark structures on the disk are known to have broad line profiles. This shows that the finding of \cite{2019ApJ...881...99M} about the linewidth is one and the same thing with the finding of \cite{2021A&A...652A..92N} and \cite{2022A&A...661L...4A} about the broadband \ha\ images. In addition, these images confirm the association of mm emission with the the network features seen in the magnetogram and in the 1600\,\AA\ band; moreover, a low brightness region appears in the  interior of a large supergranule, north of the image center, similar to the one reported by \cite{2019ApJ...877L..26L}.

\subsection{Observations of Spicules}\label{Sect:spicules}

Spicules, near the limits of instrumental resolution even at O/UV wavelengths, have not been well-studied at mm/submm-$\lambda$ because the necessary resolution has been unavailable. Nevertheless, interferometric and single disk observations during times of solar eclipses placed constraints on the mean height of the chromospheric ``limb extension" at various submm/mm wavelengths during the 1980s and 1990s. \citet{1983ApJ...264L..25L, 1986ApJ...308..448L} showed that the limb extension from 30-200$\mu$ extended well above heights expected from the hydrostatic, one-dimensional models of \citet{1981ApJS...45..635V}, a discrepancy that increased with wavelength as established by \citet{1978SoPh...59..331L, 1981ApJ...244..340H, 1983ApJ...264..660W} and later confirmed by \citet{1991ApJ...381..288R, 1992Natur.358..308L, 1992ApJ...400..692B, 1993ApJ...403..426E}. \citet{1987ApJ...320..898B} interpreted these observations in terms of a purely spicular model where the additional opacity above 1000~km was provided by a random and dynamic distribution of spicules. \citet{1992ApJ...400..692B} noted the close correspondence in height between the 3~mm limb extension and the height of H$\alpha$ spicules. \citet{1993ApJ...403..426E} pointed out that extant observations of the chromospheric limb extension was in any case consistent with an effective density scale height of 1200~km. 

ALMA provides the necessary angular resolution and time resolution to map the brightness and kinematics of solar spicules at mm/submm-$\lambda$ directly. These are important for the same reasons that they are for other problems of chromospheric physics (Section~\ref{quiet}): observations at these wavelengths can make rather direct measurements of the temperature of spicules where they are optically thick and they can impose constraints on their densities that are wholly independent of O/UV observations. 

\begin{figure}[h]
\begin{center}
\includegraphics[width=.8\textwidth]{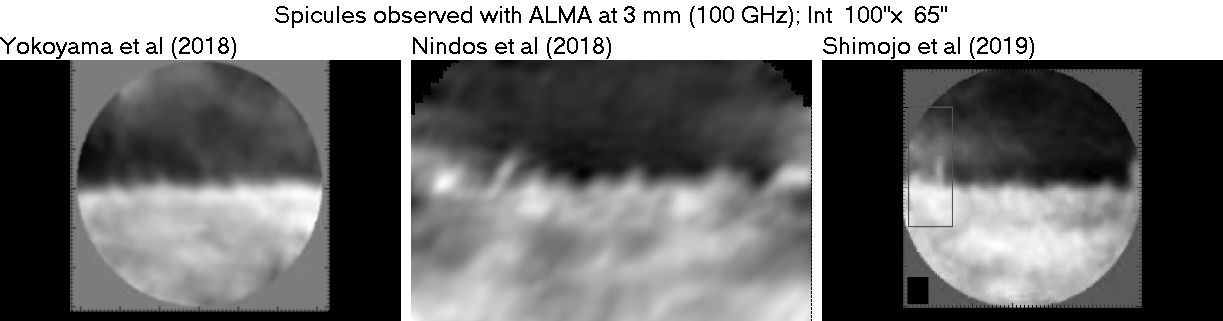}
\end{center}
\caption{The first ALMA observations of spicules at 100\,GHz (3\,mm). Images reprocessed by the authors for better visibility. Reproduced from Figure 11 of \citet{2020FrASS...7...74A}.
}
\label{ALMAspicules}
\end{figure}

As a relatively new capability, however, the number of studies of solar spicules has been small in number and scope. \citet{2018ApJ...863...96Y} reported the first observations of solar spicules (Figure~\ref{ALMAspicules} left). These were obtained at the southern limb at a wavelength of 3~mm with an angular resolution of $2.56"\times 1.6"$. These authors noted the presence of "sawtooth" irregularities on the limb and jet-like activity that appeared to correspond well to IRIS ``jet clusters". Interestingly, these limb irregularities seen at 3~mm were well-correlated with absorption observed in the SDO 171\AA\ band; i.e., the EUV emission from plasma at coronal temperatures was absorbed by foreground chromospheric structures (see also \citet{2019SoPh..294...96A}). A dynamic feature rising at 40 km-s$^{-1}$, showing a brightness excess of just 135~K was determined to have an electron density of $4.6-8.4\times 10^9$ cm$^{-3}$. 

The presence of spicules or groups of spicules was also reported by \citet{2018A&A...619L...6N}, see middle panel of Figure~\ref{ALMAspicules}, who also detected a discrete structure near the resolution of ALMA extending 15" above the limb with a life time of $\sim 10$ min and a brightness temperature of $\approx 2600$ K; these structures were identical to structures seen in simultaneous \ha\ images from the GONG network. \citet{2020ApJ...888L..28S} also reported a dynamic limb feature observed at 3~mm (Figure~\ref{ALMAspicules} right) that they interpreted as being a macrospicule. It extended to 15" above the limb with an apparent brightness temperature of 240~K. With a filling factor assumed to be 0.25 the density inferred was $\sim 10^{10}$ cm$^{-3}$. 

Clearly, work to date has only scratched the surface. Spicules are at the current limits of ALMA resolution and there are significant technical difficulties associated with the extraction of accurate brightness temperatures as evidenced by the wide range that has been reported so far. The difficulty has been highlighted by \citet{2017SoPh..292...87S} who point out that the bright solar limb causes interferometric ``overshoot", strong negative emission that is difficult to remove.  Many more observations of spicules are needed in all of the ALMA solar observing bands -- from 0.85 to 3~mm -- to more comprehensive characterize their temperature, density, and kinematics at these wavelengths as well as their relationship to UV/EUV emission and absorption.

\subsection{Visibility of spicules on the disk}\label{Sect:spicdisc}

Interestingly, no elongated structures in emission or in absorption, reminiscent of spicules appear in Figures \ref{CommQS}, \ref{targets}, \ref{ALMAnetwork} and \ref{moreALMA2}. Elongated structures (Figure~\ref{moreALMA}, left) do appear in the data used by \citet{2019ApJ...877L..26L} (Figure~\ref{moreALMA}, left), but it is not clear if these are spicules or some other kind of loops; still, movies created by the authors from data in the {\it SALSA} archive \citep{2022A&A...659A..31H} did not reveal any motions reminiscent of expanding spicules. A set of compact loops joining opposite magnetic polarities was reported by \citet{2020A&A...635A..71W}, visible in the ALMA and 171\,\AA\ images reproduced in Figure~\ref{moreALMA}, right (upper part of the field of view); the same image shows other elongated structures. We add that \citet{2021ApJ...906...82C} observed a plage region and reported an evolving elongated structure which they identified as a type II spicule.

Using values of physical parameters ($T_e,~N_e$) derived from spicule models in O/UV, it is possible to compute their brightness on the disk at mm-$\lambda$. Values from such spicule models were compiled in Table 7 of \citet{2018SoPh..293...20A}. A first look at the table shows that spicules on the disk should appear in emission, since $T_e$ values are above the mm-$\lambda$ $T_b$. A simple calculation for a 500\,km diameter vertical spicule, located 45\degr\ from the center of the disk and at a height of $\sim6000$\,km (projected distance from the spicule base of 6\arcsec) gives, for most models, brightness temperatures of the order of 5000, 2000 and 1000\,K above the background at 3, 1.26 and 0.86\,mm respectively; these values, even if they are corrected for spatial resolution of 1, 2 or 3\arcsec, by factors of 0.57, 0.33 and 0.22, respectively, predict that spicules should be prominent on the disk. Still, the ``cold model'' of  \cite{2018SoPh..293...20A} predicts excess brightness of just 140, 61 and 34\,K at the three wavelengths which, after correction for instrumental resolution; this would be compatible with the absence of spicules in disk observations. Obviously, we have a lot to learn about spicules from ALMA.

\section{Summary and future Prospects}
Five years after the first public release of the solar commissioning data, ALMA has provided valuable information that has improved our understanding of the quiet sun. Both low resolution, full disk images and high resolution interferometric imaging have had their contribution to that end.

Full disk ALMA images, of much better quality than any prior single-dish observations, made possible the comparison of the mm-$\lambda$ emission in three wavelength bands (Band 3, 3\,mm, 100\,GHz; Band 6, 1.26\,mm, 236\,GHz and Band 7, 0.86\,mm, 347\,GHz) with UV and EUV images of the solar atmosphere from the chromosphere to the corona. Practically all known solar  features were identified in such images. In particular, the chromospheric network is very prominent in the QS, and comparison with AIA images puts the formation height of the mm emission between that of the 1600\,\AA\ and 304\,\AA\ AIA bands, thus confirming that the radiation comes from the chromosphere.

Combined measurements of the brightness temperature variation from the center of the solar disk to the limb in all three ALMA bands made possible the computation of the electron temperature as a function of the optical depth and the comparison with classic atmospheric models; the ALMA results were close to model C of \cite{1993ApJ...406..319F} and far from the model C7 of \citet{2008ApJS..175..229A}. Moreover, the band 7 data showed some flattening at high optical depth, a possible indication of approach to the temperature minimum.

Interferometric ALMA observations have been published mostly for band 3, with few in band 6. The association of the mm emission with the network is confirmed by these observations, while an association with negative broadband \ha\ images was also detected. As demonstrated in Figure~\ref{moreALMA2} of this review, this is the same as the association of the emission with the \ha\ linewidth reported previously. This effect could be interpreted either in terms of \ha\ structures being transparent at mm-$\lambda$, in which case ALMA images lower-lying network elements, or in terms of opaque fibrils, directly observed by ALMA. Some cases of low brightness in the center of large supergranules were also reported.

Few ALMA observations are far from the center of the solar disk. These have allowed the measurement of the network/internetwork brightness ratio and its center-to-limb variation. Comparison with classic theoretical models showed that the network is less bright and the internetwork less dark than the model predictions.

There are only three published observations at the limb, at 3\,mm, and all show spicular structure, mostly below the image resolution. Brightness temperatures in the range of 240 to 2000\,K were reported, but these values probably suffer from the effects of insufficient spatial resolution and/or clustering of many structures. 

ALMA is a fast developing instrument and so are the solar observing modes. For the quit Sun we anticipate exciting new results as the available frequency bands increase, the spatial resolution improves, the pointing accuracy of interferometric observations is getting better, a more precise absolute calibration of full-disk images is achieved and circular polarization measurements are implemented. For details on the associated instrumentation issues we refer the reader to the article by Bastian et al. in this special Research Topic collection.

The availability of more frequency bands is important for all aspects discussed in this review. For solar atmospheric modeling in particular, we stress the importance ef extending to higher frequencies, which will take us closer to the temperature minimum. The extension to lower frequency bands is also important, as this will take us higher in the chromosphere and into the low chromosphere-corona transition region; it will also bridge the gap between ALMA and Nobeyama, as well as RATAN-600 observations. Intermediate bands (4 and 5) are no less important, as they will fill the frequency range between bands 3 and 6; we note that band 5 is already available, but no results have been published yet.  

Improved spatial resolution, which can be achieved with more extended array configurations (at the expense of less precise imaging of large-scale structure) and/or at higher frequencies, will give us a more precise view of the fine structure. This will be a great advantage for limb observation of spicules. From brightness measurements at two or more frequencies with sufficient spatial resolution, it will be straight forward to compute their electron temperature and density. The ideal situation would be to have simultaneous multi-frequency observations, but this is not possible with ALMA, at least not until a sub-array observing mode is developed. Under the current situation we have to be content with statistical analyses of images taken at different times. Improved resolution could also give information on whether features are elongated, as expected in the case of opaque fibrils, or not, as expected in the case of network structures. 

The absolute calibration of full-disk images is a rather complicated issue (see \citet{2017SoPh..292...88W}), and problems in bands 6 and 7 have been reported and remedied by \citet{2020A&A...640A..57A} and \citet{2022A&A...661L...4A}. Thus the need of a better absolute calibration of ALMA full-disk solar observations, using the moon for example, is imperative.

The pointing of interferometric solar ALMA images has a number of problems. discussed in the Bastian et al. article of this collection. A precise pointing, of the order of 1\arcsec\ or better, is important for (a) accurate measurement of the solar radius and (b) measurement of the shift of ALMA images with respect to images in other wavelengths and magnetograms, that will make possible to compute the height of the mm-$\lambda$ emission, as was done by \citet{2019SoPh..294..161A} for UV/EUV wavelengths. 

Circular polarization measurements are of primary interest for regions with high magnetic field, such as sunspots, but the signal might be too low to measure in the quiet Sun. \citet{Bogod2015} reported values of the magnetic field in the range of 40 to 200 G in the cm-$\lambda$ range for the QS, from observations with RATAN-600, which would give a polarization degree in the range of 0.22 to 1.1\% at 3\,mm. Even if this is not measurable, we will have an upper limit.

Solar ALMA work has greatly benefited from the synergy with ground and space based instruments operating in other wavelengths, such as AIA and HMI on SDO, IRIS, IBIS on the Dunn Solar Telescope and the Goode Solar Telescope (GST). As the ALMA resolution improves, this synergy will be enhanced in the near future with next-generation instruments coming online, such as the Daniel K. Inouye Solar Telescope (DKIST). It is therefore very important to plan coordinated observing campaigns.

Last but not least, the interferometric observations published so far have provided images with a small field of view and in regions mostly near the disk center; we thus lack a global view of what the mm-$\lambda$ Sun looks like at high resolution. Mosaicking of the full sun is out of question, but some large fov mosaics could be done, at the expanse of time resolution; in fact, no large mosaics have been attempted since the commissioning observations (Figure~\ref{CommQS}). The remedy is one and only: more and more solar observations with ALMA.

\section*{CONFLICT OF INTEREST STATEMENT}
The authors declare that this research was conducted in the absence of any commercial or financial relationships that could be construed as a potential conflict of interest.

\section*{AUTHOR CONTRIBUTIONS}
All authors contributed to this article.

\section*{FUNDING}
RB acknowledges the support by the Croatian Science Foundation under the project 7549 ``Millimeter and  sub-millimeter observations of the
solar chromosphere with ALMA" and by the EU Horizon 2020 project SOLARNET (824135, 2019-2023)."

\section*{Acknowledgments}
The authors thank Prof. Rob Rutten for long and interesting discussions. This work makes use of the following ALMA data: ADS/JAO.ALMA.2011.0.00020.SV, ADS/JAO.ALMA2016.1.00202.S and ADS/JAO.ALMA2016.1.00788.S. ALMA is a partnership of ESO (representing its member states), NSF (USA) and NINS (Japan), together with NRC (Canada), MOST and ASIAA (Taiwan), and KASI (Republic of Korea), in cooperation with the Republic of Chile. The Joint ALMA Observatory is operated by ESO, AUI/NRAO and NAOJ. We also used images from AIA/SDO and HMI/SDO and we are grateful to all those that operate these instruments and make their data available to the community. Finally, we thank the authors of \cite{2022ApJS..261...15K} for making their data public. The National Radio Astronomy Observatory is a facility of the National Science Foundation operated under cooperative agreement by Associated Universities, Inc.

\bibliographystyle{aasjournal} 
\bibliography{mmQS}

\begin{thebibliography}{}
\expandafter\ifx\csname natexlab\endcsname\relax\def\natexlab#1{#1}\fi
\providecommand{\url}[1]{\href{#1}{#1}}
\providecommand{\dodoi}[1]{doi:~\href{http://doi.org/#1}{\nolinkurl{#1}}}
\providecommand{\doeprint}[1]{\href{http://ascl.net/#1}{\nolinkurl{http://ascl.net/#1}}}
\providecommand{\doarXiv}[1]{\href{https://arxiv.org/abs/#1}{\nolinkurl{https://arxiv.org/abs/#1}}}

\bibitem[{{Alissandrakis}(2019)}]{2019SoPh..294..161A}
{Alissandrakis}, C.~E. 2019, \solphys, 294, 161,
  \dodoi{10.1007/s11207-019-1552-1}

\bibitem[{{Alissandrakis}(2020)}]{2020FrASS...7...74A}
---. 2020, Frontiers in Astronomy and Space Sciences, 7, 74,
  \dodoi{10.3389/fspas.2020.574460}

\bibitem[{{Alissandrakis} {et~al.}(2022){Alissandrakis}, {Bastian}, \&
  {Nindos}}]{2022A&A...661L...4A}
{Alissandrakis}, C.~E., {Bastian}, T.~S., \& {Nindos}, A. 2022, \aap, 661, L4,
  \dodoi{10.1051/0004-6361/202243774}

\bibitem[{{Alissandrakis} {et~al.}(2020){Alissandrakis}, {Nindos}, {Bastian},
  \& {Patsourakos}}]{2020A&A...640A..57A}
{Alissandrakis}, C.~E., {Nindos}, A., {Bastian}, T.~S., \& {Patsourakos}, S.
  2020, \aap, 640, A57, \dodoi{10.1051/0004-6361/202038461}

\bibitem[{{Alissandrakis} {et~al.}(2017){Alissandrakis}, {Patsourakos},
  {Nindos}, \& {Bastian}}]{2017A&A...605A..78A}
{Alissandrakis}, C.~E., {Patsourakos}, S., {Nindos}, A., \& {Bastian}, T.~S.
  2017, \aap, 605, A78, \dodoi{10.1051/0004-6361/201730953}

\bibitem[{{Alissandrakis} \& {Valentino}(2019)}]{2019SoPh..294...96A}
{Alissandrakis}, C.~E., \& {Valentino}, A. 2019, \solphys, 294, 96,
  \dodoi{10.1007/s11207-019-1486-7}

\bibitem[{{Alissandrakis} {et~al.}(2018){Alissandrakis}, {Vial}, {Koukras},
  {Buchlin}, \& {Chane-Yook}}]{2018SoPh..293...20A}
{Alissandrakis}, C.~E., {Vial}, J.~C., {Koukras}, A., {Buchlin}, E., \&
  {Chane-Yook}, M. 2018, \solphys, 293, 20, \dodoi{10.1007/s11207-018-1242-4}

\bibitem[{{Avrett} \& {Loeser}(2008)}]{2008ApJS..175..229A}
{Avrett}, E.~H., \& {Loeser}, R. 2008, \apjs, 175, 229, \dodoi{10.1086/523671}

\bibitem[{{Bastian} {et~al.}(2017){Bastian}, {Chintzoglou}, {De Pontieu},
  {Shimojo}, {Schmit}, {Leenaarts}, \& {Loukitcheva}}]{2017ApJ...845L..19B}
{Bastian}, T.~S., {Chintzoglou}, G., {De Pontieu}, B., {et~al.} 2017, \apjl,
  845, L19, \dodoi{10.3847/2041-8213/aa844c}

\bibitem[{{Bastian} {et~al.}(1993){Bastian}, {Ewell}, \&
  {Zirin}}]{1993ApJ...415..364B}
{Bastian}, T.~S., {Ewell}, M.~W., J., \& {Zirin}, H. 1993, \apj, 415, 364,
  \dodoi{10.1086/173170}

\bibitem[{{Beckers}(1968)}]{1968SoPh....3..367B}
{Beckers}, J.~M. 1968, \solphys, 3, 367, \dodoi{10.1007/BF00171614}

\bibitem[{{Beckers}(1972)}]{1972ARA&A..10...73B}
---. 1972, \araa, 10, 73, \dodoi{10.1146/annurev.aa.10.090172.000445}

\bibitem[{{Belkora} {et~al.}(1992){Belkora}, {Hurford}, {Gary}, \&
  {Woody}}]{1992ApJ...400..692B}
{Belkora}, L., {Hurford}, G.~J., {Gary}, D.~E., \& {Woody}, D.~P. 1992, \apj,
  400, 692, \dodoi{10.1086/172031}

\bibitem[{{Benz}(2009)}]{Benz2009}
{Benz}, A.~O. 2009, Landolt B{\"o}rnstein, 4116,
  \dodoi{10.1007/978-3-540-88055-4_5}

\bibitem[{Bogod {et~al.}(2015)Bogod, Alissandrakis, Kaltman, \&
  Tokhchukova}]{Bogod2015}
Bogod, V.~M., Alissandrakis, C.~E., Kaltman, T.~I., \& Tokhchukova, S.~K. 2015,
  \solphys, 290, 7, \dodoi{10.1007/s11207-014-0526-6}

\bibitem[{{Bohlin} {et~al.}(1975){Bohlin}, {Vogel}, {Purcell}, {Sheeley},
  {Tousey}, \& {Vanhoosier}}]{1975ApJ...197L.133B}
{Bohlin}, J.~D., {Vogel}, S.~N., {Purcell}, J.~D., {et~al.} 1975, \apjl, 197,
  L133, \dodoi{10.1086/181794}

\bibitem[{{Bose} {et~al.}(2019){Bose}, {Henriques}, {Joshi}, \& {Rouppe van der
  Voort}}]{2019A&A...631L...5B}
{Bose}, S., {Henriques}, V. M.~J., {Joshi}, J., \& {Rouppe van der Voort}, L.
  2019, \aap, 631, L5, \dodoi{10.1051/0004-6361/201936617}

\bibitem[{{Bose} {et~al.}(2021){Bose}, {Joshi}, {Henriques}, \& {Rouppe van der
  Voort}}]{2021A&A...647A.147B}
{Bose}, S., {Joshi}, J., {Henriques}, V. M.~J., \& {Rouppe van der Voort}, L.
  2021, \aap, 647, A147, \dodoi{10.1051/0004-6361/202040014}

\bibitem[{{Braj{\v s}a} {et~al.}(2007{\natexlab{a}}){Braj{\v s}a}, {Benz},
  {Temmer}, {Jurdana-{\v S}epi{\'c}}, {{\v S}aina}, \&
  {W{\"o}hl}}]{Brajsa2007a}
{Braj{\v s}a}, R., {Benz}, A.~O., {Temmer}, M., {et~al.} 2007{\natexlab{a}},
  \solphys, 245, 167, \dodoi{10.1007/s11207-007-9008-4}

\bibitem[{{Braj{\v s}a} {et~al.}(2007{\natexlab{b}}){Braj{\v s}a}, {Benz},
  {Temmer}, {Jurdana-{\v S}epi{\'c}}, {{\v S}aina}, {W{\"o}hl}, \& {Ru{\v
  z}djak}}]{Brajsa2007b}
---. 2007{\natexlab{b}}, Central European Astrophysical Bulletin, 31, 219

\bibitem[{{Braj{\v{s}}a} {et~al.}(2021){Braj{\v{s}}a}, {Skoki{\'c}}, {Sudar},
  {Benz}, {Krucker}, {Ludwig}, {Saar}, \& {Selhorst}}]{2021A&A...651A...6B}
{Braj{\v{s}}a}, R., {Skoki{\'c}}, I., {Sudar}, D., {et~al.} 2021, \aap, 651,
  A6, \dodoi{10.1051/0004-6361/201936231}

\bibitem[{{Braj{\v{s}}a} {et~al.}(2018){Braj{\v{s}}a}, {Sudar}, {Benz},
  {Skoki{\'c}}, {B{\'a}rta}, {De Pontieu}, {Kim}, {Kobelski}, {Kuhar},
  {Shimojo}, {Wedemeyer}, {White}, {Yagoubov}, \& {Yan}}]{2018A&A...613A..17B}
{Braj{\v{s}}a}, R., {Sudar}, D., {Benz}, A.~O., {et~al.} 2018, \aap, 613, A17,
  \dodoi{10.1051/0004-6361/201730656}

\bibitem[{{Braun} \& {Lindsey}(1987)}]{1987ApJ...320..898B}
{Braun}, D., \& {Lindsey}, C. 1987, \apj, 320, 898, \dodoi{10.1086/165607}

\bibitem[{{Buhl} \& {Tlamicha}(1970)}]{1970A&A.....5..102B}
{Buhl}, D., \& {Tlamicha}, A. 1970, \aap, 5, 102

\bibitem[{{Carlsson} {et~al.}(2019){Carlsson}, {De Pontieu}, \&
  {Hansteen}}]{2019ARA&A..57..189C}
{Carlsson}, M., {De Pontieu}, B., \& {Hansteen}, V.~H. 2019, \araa, 57, 189,
  \dodoi{10.1146/annurev-astro-081817-052044}

\bibitem[{{Carlsson} {et~al.}(2016){Carlsson}, {Hansteen}, {Gudiksen},
  {Leenaarts}, \& {De Pontieu}}]{2016A&A...585A...4C}
{Carlsson}, M., {Hansteen}, V.~H., {Gudiksen}, B.~V., {Leenaarts}, J., \& {De
  Pontieu}, B. 2016, \aap, 585, A4, \dodoi{10.1051/0004-6361/201527226}

\bibitem[{{Chintzoglou} {et~al.}(2021){Chintzoglou}, {De Pontieu},
  {Mart{\'\i}nez-Sykora}, {Hansteen}, {de la Cruz Rodr{\'\i}guez},
  {Szydlarski}, {Jafarzadeh}, {Wedemeyer}, {Bastian}, \& {Sainz
  Dalda}}]{2021ApJ...906...82C}
{Chintzoglou}, G., {De Pontieu}, B., {Mart{\'\i}nez-Sykora}, J., {et~al.} 2021,
  \apj, 906, 82, \dodoi{10.3847/1538-4357/abc9b1}

\bibitem[{{De Pontieu} {et~al.}(2004){De Pontieu}, {Erd{\'e}lyi}, \&
  {James}}]{2004Natur.430..536D}
{De Pontieu}, B., {Erd{\'e}lyi}, R., \& {James}, S.~P. 2004, \nat, 430, 536,
  \dodoi{10.1038/nature02749}

\bibitem[{{de Pontieu} {et~al.}(2007){de Pontieu}, {McIntosh}, {Hansteen},
  {Carlsson}, {Schrijver}, {Tarbell}, {Title}, {Shine}, {Suematsu}, {Tsuneta},
  {Katsukawa}, {Ichimoto}, {Shimizu}, \& {Nagata}}]{2007PASJ...59S.655D}
{de Pontieu}, B., {McIntosh}, S., {Hansteen}, V.~H., {et~al.} 2007, \pasj, 59,
  S655, \dodoi{10.1093/pasj/59.sp3.S655}

\bibitem[{{Eklund} {et~al.}(2021){Eklund}, {Wedemeyer}, {Szydlarski}, \&
  {Jafarzadeh}}]{2021A&A...656A..68E}
{Eklund}, H., {Wedemeyer}, S., {Szydlarski}, M., \& {Jafarzadeh}, S. 2021,
  \aap, 656, A68, \dodoi{10.1051/0004-6361/202140972}

\bibitem[{{Ewell} {et~al.}(1993){Ewell}, {Zirin}, {Jensen}, \&
  {Bastian}}]{1993ApJ...403..426E}
{Ewell}, M.~W., J., {Zirin}, H., {Jensen}, J.~B., \& {Bastian}, T.~S. 1993,
  \apj, 403, 426, \dodoi{10.1086/172213}

\bibitem[{{Fontenla} {et~al.}(1993){Fontenla}, {Avrett}, \&
  {Loeser}}]{1993ApJ...406..319F}
{Fontenla}, J.~M., {Avrett}, E.~H., \& {Loeser}, R. 1993, \apj, 406, 319,
  \dodoi{10.1086/172443}

\bibitem[{{Gary}(1996)}]{Gary1996}
{Gary}, D.~E. 1996, in Astronomical Society of the Pacific Conference Series,
  Vol.~93, Radio Emission from the Stars and the Sun, ed. A.~R. {Taylor} \&
  J.~M. {Paredes}, 387

\bibitem[{{Gary}(2001)}]{2001SoPh..203...71G}
{Gary}, G.~A. 2001, \solphys, 203, 71, \dodoi{10.1023/A:1012722021820}

\bibitem[{{Henriques} {et~al.}(2022){Henriques}, {Jafarzadeh}, {Guevara
  G{\'o}mez}, {Eklund}, {Wedemeyer}, {Szydlarski}, {Haugan}, \&
  {Mohan}}]{2022A&A...659A..31H}
{Henriques}, V. M.~J., {Jafarzadeh}, S., {Guevara G{\'o}mez}, J.~C., {et~al.}
  2022, \aap, 659, A31, \dodoi{10.1051/0004-6361/202142291}

\bibitem[{{Horne} {et~al.}(1981){Horne}, {Hurford}, {Zirin}, \& {de
  Graauw}}]{1981ApJ...244..340H}
{Horne}, K., {Hurford}, G.~J., {Zirin}, H., \& {de Graauw}, T. 1981, \apj, 244,
  340, \dodoi{10.1086/158711}

\bibitem[{{Iwai} {et~al.}(2017){Iwai}, {Shimojo}, {Asayama}, {Minamidani},
  {White}, {Bastian}, \& {Saito}}]{2017SoPh..292...22I}
{Iwai}, K., {Shimojo}, M., {Asayama}, S., {et~al.} 2017, \solphys, 292, 22,
  \dodoi{10.1007/s11207-016-1044-5}

\bibitem[{{Kleint} \& {Panos}(2022)}]{2022A&A...657A.132K}
{Kleint}, L., \& {Panos}, B. 2022, \aap, 657, A132,
  \dodoi{10.1051/0004-6361/202142235}

\bibitem[{{Kobelski} {et~al.}(2022){Kobelski}, {Tarr}, {Jaeggli}, {Luber},
  {Warren}, \& {Savage}}]{2022ApJS..261...15K}
{Kobelski}, A.~R., {Tarr}, L.~A., {Jaeggli}, S.~A., {et~al.} 2022, \apjs, 261,
  15, \dodoi{10.3847/1538-4365/ac6b3b}

\bibitem[{{Kundu}(1970)}]{1970SoPh...13..348K}
{Kundu}, M.~R. 1970, \solphys, 13, 348, \dodoi{10.1007/BF00153556}

\bibitem[{{Kundu}(1971)}]{1971SoPh...21..130K}
---. 1971, \solphys, 21, 130, \dodoi{10.1007/BF00155783}

\bibitem[{{Kundu}(1972)}]{1972SoPh...25..108K}
---. 1972, \solphys, 25, 108, \dodoi{10.1007/BF00155749}

\bibitem[{{Kundu} \& {McCullough}(1972)}]{1972SoPh...24..133K}
{Kundu}, M.~R., \& {McCullough}, T.~P. 1972, \solphys, 24, 133,
  \dodoi{10.1007/BF00231091}

\bibitem[{{Labrum} {et~al.}(1978){Labrum}, {Archer}, \&
  {Smith}}]{1978SoPh...59..331L}
{Labrum}, N.~R., {Archer}, J.~W., \& {Smith}, C.~J. 1978, \solphys, 59, 331,
  \dodoi{10.1007/BF00951837}

\bibitem[{{Langangen} {et~al.}(2008){Langangen}, {De Pontieu}, {Carlsson},
  {Hansteen}, {Cauzzi}, \& {Reardon}}]{2008ApJ...679L.167L}
{Langangen}, {\O}., {De Pontieu}, B., {Carlsson}, M., {et~al.} 2008, \apjl,
  679, L167, \dodoi{10.1086/589442}

\bibitem[{{Lantos} \& {Kundu}(1972)}]{1972A&A....21..119L}
{Lantos}, P., \& {Kundu}, M.~R. 1972, \aap, 21, 119

\bibitem[{{Lindsey} {et~al.}(1983){Lindsey}, {Becklin}, {Jefferies}, {Orrall},
  {Werner}, \& {Gatley}}]{1983ApJ...264L..25L}
{Lindsey}, C., {Becklin}, E.~E., {Jefferies}, J.~T., {et~al.} 1983, \apjl, 264,
  L25, \dodoi{10.1086/183938}

\bibitem[{{Lindsey} {et~al.}(1984){Lindsey}, {Becklin}, {Jefferies}, {Orrall},
  {Werner}, \& {Gatley}}]{1984ApJ...281..862L}
---. 1984, \apj, 281, 862, \dodoi{10.1086/162166}

\bibitem[{{Lindsey} {et~al.}(1986){Lindsey}, {Becklin}, {Orrall}, {Werner},
  {Jefferies}, \& {Gatley}}]{1986ApJ...308..448L}
{Lindsey}, C., {Becklin}, E.~E., {Orrall}, F.~Q., {et~al.} 1986, \apj, 308,
  448, \dodoi{10.1086/164515}

\bibitem[{{Lindsey} {et~al.}(1992){Lindsey}, {Jefferies}, {Clark}, {Harrison},
  {Carter}, {Watt}, {Becklin}, {Roellig}, {Braun}, \&
  {Naylor}}]{1992Natur.358..308L}
{Lindsey}, C., {Jefferies}, J.~T., {Clark}, T.~A., {et~al.} 1992, \nat, 358,
  308, \dodoi{10.1038/358308a0}

\bibitem[{{Lindsey} {et~al.}(1990){Lindsey}, {Yee}, {Roellig}, {Hills},
  {Brock}, {Duncan}, {Watt}, {Webster}, \& {Jefferies}}]{1990ApJ...353L..53L}
{Lindsey}, C.~A., {Yee}, S., {Roellig}, T.~L., {et~al.} 1990, \apjl, 353, L53,
  \dodoi{10.1086/185706}

\bibitem[{{Linsky}(1973)}]{1973SoPh...28..409L}
{Linsky}, J.~L. 1973, \solphys, 28, 409, \dodoi{10.1007/BF00152312}

\bibitem[{{Loukitcheva} {et~al.}(2004{\natexlab{a}}){Loukitcheva}, {Solanki},
  {Carlsson}, \& {Stein}}]{2004A&A...419..747L}
{Loukitcheva}, M., {Solanki}, S.~K., {Carlsson}, M., \& {Stein}, R.~F.
  2004{\natexlab{a}}, \aap, 419, 747, \dodoi{10.1051/0004-6361:20034159}

\bibitem[{{Loukitcheva} {et~al.}(2004{\natexlab{b}}){Loukitcheva}, {Solanki},
  {Carlsson}, \& {Stein}}]{Loukitcheva2004}
---. 2004{\natexlab{b}}, \aap, 419, 747, \dodoi{10.1051/0004-6361:20034159}

\bibitem[{{Loukitcheva} {et~al.}(2015){Loukitcheva}, {Solanki}, {Carlsson}, \&
  {White}}]{Loukitcheva2015}
{Loukitcheva}, M., {Solanki}, S.~K., {Carlsson}, M., \& {White}, S.~M. 2015,
  \aap, 575, A15, \dodoi{10.1051/0004-6361/201425238}

\bibitem[{{Loukitcheva} {et~al.}(2006){Loukitcheva}, {Solanki}, \&
  {White}}]{2006A&A...456..713L}
{Loukitcheva}, M., {Solanki}, S.~K., \& {White}, S. 2006, \aap, 456, 713,
  \dodoi{10.1051/0004-6361:20053171}

\bibitem[{Loukitcheva {et~al.}(2019)Loukitcheva, White, \&
  Solanki}]{2019ApJ...877L..26L}
Loukitcheva, M.~A., White, S.~M., \& Solanki, S.~K. 2019, \apjl, 877, L26,
  \dodoi{10.3847/2041-8213/ab2191}

\bibitem[{{Macris}(1957)}]{1957AnAp...20..179M}
{Macris}, C.~J. 1957, Annales d'Astrophysique, 20, 179

\bibitem[{{Mart{\'\i}nez-Sykora} {et~al.}(2020){Mart{\'\i}nez-Sykora}, {De
  Pontieu}, {de la Cruz Rodriguez}, \& {Chintzoglou}}]{2020ApJ...891L...8M}
{Mart{\'\i}nez-Sykora}, J., {De Pontieu}, B., {de la Cruz Rodriguez}, J., \&
  {Chintzoglou}, G. 2020, \apjl, 891, L8, \dodoi{10.3847/2041-8213/ab75ac}

\bibitem[{{Mart{\'\i}nez-Sykora} {et~al.}(2017){Mart{\'\i}nez-Sykora}, {De
  Pontieu}, {Hansteen}, {Rouppe van der Voort}, {Carlsson}, \&
  {Pereira}}]{2017Sci...356.1269M}
{Mart{\'\i}nez-Sykora}, J., {De Pontieu}, B., {Hansteen}, V.~H., {et~al.} 2017,
  Science, 356, 1269, \dodoi{10.1126/science.aah5412}

\bibitem[{{Menezes} {et~al.}(2022){Menezes}, {Selhorst}, {Gim{\'e}nez de
  Castro}, \& {Valio}}]{2022MNRAS.511..877M}
{Menezes}, F., {Selhorst}, C.~L., {Gim{\'e}nez de Castro}, C.~G., \& {Valio},
  A. 2022, \mnras, 511, 877, \dodoi{10.1093/mnras/stab3501}

\bibitem[{{Molnar} {et~al.}(2019){Molnar}, {Reardon}, {Chai}, {Gary},
  {Uitenbroek}, {Cauzzi}, \& {Cranmer}}]{2019ApJ...881...99M}
{Molnar}, M.~E., {Reardon}, K.~P., {Chai}, Y., {et~al.} 2019, \apj, 881, 99,
  \dodoi{10.3847/1538-4357/ab2ba3}

\bibitem[{{Nindos} {et~al.}(2021){Nindos}, {Patsourakos}, {Alissandrakis}, \&
  {Bastian}}]{2021A&A...652A..92N}
{Nindos}, A., {Patsourakos}, S., {Alissandrakis}, C.~E., \& {Bastian}, T.~S.
  2021, \aap, 652, A92, \dodoi{10.1051/0004-6361/202141241}

\bibitem[{{Nindos} {et~al.}(2018){Nindos}, {Alissandrakis}, {Bastian},
  {Patsourakos}, {De Pontieu}, {Warren}, {Ayres}, {Hudson}, {Shimizu}, {Vial},
  {Wedemeyer}, \& {Yurchyshyn}}]{2018A&A...619L...6N}
{Nindos}, A., {Alissandrakis}, C.~E., {Bastian}, T.~S., {et~al.} 2018, \aap,
  619, L6, \dodoi{10.1051/0004-6361/201834113}

\bibitem[{{O'Dwyer} {et~al.}(2010){O'Dwyer}, {Del Zanna}, {Mason}, {Weber}, \&
  {Tripathi}}]{2010A&A...521A..21O}
{O'Dwyer}, B., {Del Zanna}, G., {Mason}, H.~E., {Weber}, M.~A., \& {Tripathi},
  D. 2010, \aap, 521, A21, \dodoi{10.1051/0004-6361/201014872}

\bibitem[{{Parker}(1988)}]{1988ApJ...330..474P}
{Parker}, E.~N. 1988, \apj, 330, 474, \dodoi{10.1086/166485}

\bibitem[{{Pikel'Ner}(1969)}]{1969AZh....46..328P}
{Pikel'Ner}, S.~B. 1969, \azh, 46, 328

\bibitem[{{Purkhart} \& {Veronig}(2022)}]{2022A&A...661A.149P}
{Purkhart}, S., \& {Veronig}, A.~M. 2022, \aap, 661, A149,
  \dodoi{10.1051/0004-6361/202243234}

\bibitem[{{Reeves} {et~al.}(1974){Reeves}, {Foukal}, {Huber}, {Noyes},
  {Schmahl}, {Timothy}, {Vernazza}, \& {Withbroe}}]{1974ApJ...188L..27R}
{Reeves}, E.~M., {Foukal}, P.~V., {Huber}, M.~C.~E., {et~al.} 1974, \apjl, 188,
  L27, \dodoi{10.1086/181423}

\bibitem[{{Roellig} {et~al.}(1991){Roellig}, {Becklin}, {Jefferies}, {Kopp},
  {Lindsey}, {Orrall}, \& {Werner}}]{1991ApJ...381..288R}
{Roellig}, T.~L., {Becklin}, E.~E., {Jefferies}, J.~T., {et~al.} 1991, \apj,
  381, 288, \dodoi{10.1086/170650}

\bibitem[{{Rouppe van der Voort} {et~al.}(2009){Rouppe van der Voort},
  {Leenaarts}, {de Pontieu}, {Carlsson}, \& {Vissers}}]{2009ApJ...705..272R}
{Rouppe van der Voort}, L., {Leenaarts}, J., {de Pontieu}, B., {Carlsson}, M.,
  \& {Vissers}, G. 2009, \apj, 705, 272, \dodoi{10.1088/0004-637X/705/1/272}

\bibitem[{{Rutten}(2017)}]{2017A&A...598A..89R}
{Rutten}, R.~J. 2017, \aap, 598, A89, \dodoi{10.1051/0004-6361/201629238}

\bibitem[{{Rutten} \& {Rouppe van der Voort}(2017)}]{2017A&A...597A.138R}
{Rutten}, R.~J., \& {Rouppe van der Voort}, L.~H.~M. 2017, \aap, 597, A138,
  \dodoi{10.1051/0004-6361/201527560}

\bibitem[{{Rutten} {et~al.}(2019){Rutten}, {Rouppe van der Voort}, \& {De
  Pontieu}}]{2019A&A...632A..96R}
{Rutten}, R.~J., {Rouppe van der Voort}, L. H.~M., \& {De Pontieu}, B. 2019,
  \aap, 632, A96, \dodoi{10.1051/0004-6361/201936113}

\bibitem[{{Samanta} {et~al.}(2019){Samanta}, {Tian}, {Yurchyshyn}, {Peter},
  {Cao}, {Sterling}, {Erd{\'e}lyi}, {Ahn}, {Feng}, {Utz}, {Banerjee}, \&
  {Chen}}]{2019Sci...366..890S}
{Samanta}, T., {Tian}, H., {Yurchyshyn}, V., {et~al.} 2019, Science, 366, 890,
  \dodoi{10.1126/science.aaw2796}

\bibitem[{{Saqri} {et~al.}(2022){Saqri}, {Veronig}, {Warmuth}, {Dickson},
  {Battaglia}, {Podladchikova}, {Xiao}, {Battaglia}, {Hurford}, \&
  {Krucker}}]{2022A&A...659A..52S}
{Saqri}, J., {Veronig}, A.~M., {Warmuth}, A., {et~al.} 2022, \aap, 659, A52,
  \dodoi{10.1051/0004-6361/202142373}

\bibitem[{{Selhorst} {et~al.}(2005){Selhorst}, {Silva}, \&
  {Costa}}]{2005A&A...433..365S}
{Selhorst}, C.~L., {Silva}, A.~V.~R., \& {Costa}, J.~E.~R. 2005, \aap, 433,
  365, \dodoi{10.1051/0004-6361:20042043}

\bibitem[{{Selhorst} {et~al.}(2019){Selhorst}, {Sim{\~o}es}, {Braj{\v{s}}a},
  {Valio}, {Gim{\'e}nez de Castro}, {Costa}, {Menezes}, {Rozelot}, {Hales},
  {Iwai}, \& {White}}]{2019ApJ...871...45S}
{Selhorst}, C.~L., {Sim{\~o}es}, P. J.~A., {Braj{\v{s}}a}, R., {et~al.} 2019,
  \apj, 871, 45, \dodoi{10.3847/1538-4357/aaf4f2}

\bibitem[{{Shibasaki} {et~al.}(2011){Shibasaki}, {Alissandrakis}, \&
  {Pohjolainen}}]{2011SoPh..273..309S}
{Shibasaki}, K., {Alissandrakis}, C.~E., \& {Pohjolainen}, S. 2011, \solphys,
  273, 309, \dodoi{10.1007/s11207-011-9788-4}

\bibitem[{{Shimojo} {et~al.}(2017){Shimojo}, {Bastian}, {Hales}, {White},
  {Iwai}, {Hills}, {Hirota}, {Phillips}, {Sawada}, {Yagoubov}, {Siringo},
  {Asayama}, {Sugimoto}, {Braj{\v{s}}a}, {Skoki{\'c}}, {B{\'a}rta}, {Kim}, {de
  Gregorio-Monsalvo}, {Corder}, {Hudson}, {Wedemeyer}, {Gary}, {De Pontieu},
  {Loukitcheva}, {Fleishman}, {Chen}, {Kobelski}, \&
  {Yan}}]{2017SoPh..292...87S}
{Shimojo}, M., {Bastian}, T.~S., {Hales}, A.~S., {et~al.} 2017, \solphys, 292,
  87, \dodoi{10.1007/s11207-017-1095-2}

\bibitem[{{Shimojo} {et~al.}(2020){Shimojo}, {Kawate}, {Okamoto}, {Yokoyama},
  {Narukage}, {Sakao}, {Iwai}, {Fleishman}, \& {Shibata}}]{2020ApJ...888L..28S}
{Shimojo}, M., {Kawate}, T., {Okamoto}, T.~J., {et~al.} 2020, \apjl, 888, L28,
  \dodoi{10.3847/2041-8213/ab62a5}

\bibitem[{{Shokri} {et~al.}(2022){Shokri}, {Alipour}, {Safari}, {Kayshap},
  {Podladchikova}, {Nigro}, \& {Tripathi}}]{2022ApJ...926...42S}
{Shokri}, Z., {Alipour}, N., {Safari}, H., {et~al.} 2022, \apj, 926, 42,
  \dodoi{10.3847/1538-4357/ac4265}

\bibitem[{{Sudar} {et~al.}(2019){Sudar}, {Braj{\v{s}}a}, {Skoki{\'c}}, \&
  {Benz}}]{2019SoPh..294..163S}
{Sudar}, D., {Braj{\v{s}}a}, R., {Skoki{\'c}}, I., \& {Benz}, A.~O. 2019,
  \solphys, 294, 163, \dodoi{10.1007/s11207-019-1556-x}

\bibitem[{Tsiropoula {et~al.}(2012)Tsiropoula, Tziotziou, Kontogiannis,
  Madjarska, Doyle, \& Suematsu}]{Tsiropoula2012}
Tsiropoula, G., Tziotziou, K., Kontogiannis, I., {et~al.} 2012, \ssr, 169, 181,
  \dodoi{10.1007/s11214-012-9920-2}

\bibitem[{{Vernazza} {et~al.}(1981){Vernazza}, {Avrett}, \&
  {Loeser}}]{1981ApJS...45..635V}
{Vernazza}, J.~E., {Avrett}, E.~H., \& {Loeser}, R. 1981, \apjs, 45, 635,
  \dodoi{10.1086/190731}

\bibitem[{{Wannier} {et~al.}(1983){Wannier}, {Hurford}, \&
  {Seielstad}}]{1983ApJ...264..660W}
{Wannier}, P.~G., {Hurford}, G.~J., \& {Seielstad}, G.~A. 1983, \apj, 264, 660,
  \dodoi{10.1086/160639}

\bibitem[{{Wedemeyer} {et~al.}(2016){Wedemeyer}, {Bastian}, {Braj{\v{s}}a},
  {Hudson}, {Fleishman}, {Loukitcheva}, {Fleck}, {Kontar}, {De Pontieu},
  {Yagoubov}, {Tiwari}, {Soler}, {Black}, {Antolin}, {Scullion}, {Gun{\'a}r},
  {Labrosse}, {Ludwig}, {Benz}, {White}, {Hauschildt}, {Doyle}, {Nakariakov},
  {Ayres}, {Heinzel}, {Karlicky}, {Van Doorsselaere}, {Gary}, {Alissandrakis},
  {Nindos}, {Solanki}, {Rouppe van der Voort}, {Shimojo}, {Kato},
  {Zaqarashvili}, {Perez}, {Selhorst}, \& {Barta}}]{2016SSRv..200....1W}
{Wedemeyer}, S., {Bastian}, T., {Braj{\v{s}}a}, R., {et~al.} 2016, \ssr, 200,
  1, \dodoi{10.1007/s11214-015-0229-9}

\bibitem[{{Wedemeyer} {et~al.}(2020){Wedemeyer}, {Szydlarski}, {Jafarzadeh},
  {Eklund}, {Guevara Gomez}, {Bastian}, {Fleck}, {de la Cruz Rodriguez},
  {Rodger}, \& {Carlsson}}]{2020A&A...635A..71W}
{Wedemeyer}, S., {Szydlarski}, M., {Jafarzadeh}, S., {et~al.} 2020, \aap, 635,
  A71, \dodoi{10.1051/0004-6361/201937122}

\bibitem[{{White} {et~al.}(2006){White}, {Loukitcheva}, \&
  {Solanki}}]{2006A&A...456..697W}
{White}, S.~M., {Loukitcheva}, M., \& {Solanki}, S.~K. 2006, \aap, 456, 697,
  \dodoi{10.1051/0004-6361:20052854}

\bibitem[{{White} {et~al.}(2017){White}, {Iwai}, {Phillips}, {Hills}, {Hirota},
  {Yagoubov}, {Siringo}, {Shimojo}, {Bastian}, {Hales}, {Sawada}, {Asayama},
  {Sugimoto}, {Marson}, {Kawasaki}, {Muller}, {Nakazato}, {Sugimoto},
  {Braj{\v{s}}a}, {Skoki{\'c}}, {B{\'a}rta}, {Kim}, {Remijan}, {de Gregorio},
  {Corder}, {Hudson}, {Loukitcheva}, {Chen}, {De Pontieu}, {Fleishmann},
  {Gary}, {Kobelski}, {Wedemeyer}, \& {Yan}}]{2017SoPh..292...88W}
{White}, S.~M., {Iwai}, K., {Phillips}, N.~M., {et~al.} 2017, \solphys, 292,
  88, \dodoi{10.1007/s11207-017-1123-2}

\bibitem[{{Wootten} \& {Thompson}(2009)}]{2009IEEEP..97.1463W}
{Wootten}, A., \& {Thompson}, A.~R. 2009, IEEE Proceedings, 97, 1463,
  \dodoi{10.1109/JPROC.2009.2020572}

\bibitem[{{Yokoyama} {et~al.}(2018){Yokoyama}, {Shimojo}, {Okamoto}, \&
  {Iijima}}]{2018ApJ...863...96Y}
{Yokoyama}, T., {Shimojo}, M., {Okamoto}, T.~J., \& {Iijima}, H. 2018, \apj,
  863, 96, \dodoi{10.3847/1538-4357/aad27e}

\end{thebibliography}

\end{document}